\documentclass[preprint,11pt]{elsarticle}

\usepackage{amssymb}
\usepackage{amsmath}

\usepackage[hidelinks]{hyperref}

\usepackage{tablefootnote}
\usepackage[section]{placeins}
\usepackage[english]{babel}

\journal{}

\begin{document}

\begin{frontmatter}



\title{Library transfer between distinct Laser-Induced Breakdown Spectroscopy systems with shared standards}


\author[inst1,inst2]{Vrábel, J.\corref{cor1}}
\ead{jakub.vrabel@ceitec.vutbr.cz}
\author[inst1,inst2]{Képeš, E.}
\author[inst1]{Nedělník, P.}
\author[inst1,inst2]{Buday, J.}
\author[inst3]{Cempírek, J.}
\author[inst1,inst2]{Pořízka, P.\corref{cor1}}
\ead{pavel.porizka@ceitec.vutbr.cz}
\cortext[cor1]{}

\author[inst1,inst2]{Kaiser, J.}

\affiliation[inst1]{organization={CEITEC, Brno University of Technology},
            addressline={Purkyňova~123}, 
            city={Brno},
            postcode={612 00}, 
            country={Czech Republic}}

\affiliation[inst2]{organization={Institute of Physical Engineering, Brno University of Technology},
            addressline={Technická~2}, 
            city={Brno},
            postcode={616 00}, 
            country={Czech Republic}}

\affiliation[inst3]{organization={Department of Geological Sciences, Faculty of Science, Masaryk University},
            addressline={Kotlářská~2}, 
            city={Brno},
            postcode={602 00}, 
            country={Czech Republic}}
            
\begin{abstract}
The mutual incompatibility of distinct spectroscopic systems is among the most limiting factors in Laser-Induced Breakdown Spectroscopy (LIBS). The cost related to setting up a new LIBS system is increased, as its extensive calibration is required. Solving the problem would enable inter-laboratory reference measurements and shared spectral libraries, which are fundamental for other spectroscopic techniques. In this work, we study a simplified version of this challenge where LIBS systems differ only in used spectrometers and collection optics but share all other parts of the apparatus, and collect spectra simultaneously from the same plasma plume. Extensive datasets measured as hyperspectral images of heterogeneous specimens are used to train machine learning models that can transfer spectra between systems. The transfer is realized by a pipeline that consists of a variational autoencoder (VAE) and a fully-connected artificial neural network (ANN). In the first step, we obtain a latent representation of the spectra which were measured on the Primary system (by using the VAE). In the second step, we map spectra from the Secondary system to corresponding locations in the latent space (by the ANN). Finally, Secondary system spectra are reconstructed from the latent space to the space of the Primary system. The transfer is evaluated by several figures of merit (Euclidean and cosine distances, both spatially resolved; k-means clustering of transferred spectra). The methodology is compared to several baseline approaches.  
\end{abstract}


\end{frontmatter}


\section{Introduction}
\label{sec:Intro}
\noindent
Laser-induced breakdown spectroscopy (LIBS) is an optical emission spectroscopic technique with rapid real-time sensing capabilities, providing reliable qualitative, and semi-quantitative analysis. LIBS is gaining traction as a laboratory technique, e.g., for producing large (megapixel) high-resolution (tens of \textmu m lateral resolution) hyperspectral images~\cite{limbeck2021methodology, MOTTOROS2020329}, with a high relevance in biological~\cite{MODLITBOVA2021}, geological~\cite{nardecchia2020detection}, and industrial settings~\cite{jolivet2020quantitative}. LIBS is often preferred for in-situ analyses owing to its stand-off capabilities, robust instrumentation, and low demands on sample preparation. Consequently, LIBS is being used for extraterrestrial exploration; namely the ChemCam device on the Curiosity rover~\cite{Wiens2012} and SuperCam on the Perseverance rover~\cite{Wiens2020}.

As a trade-off to its instrumental robustness, LIBS exhibits a high sensitivity to changes e.g. in the analyzed target’s topography~\cite{SALAJKOVA2021}, and matrix~\cite{TAKAHASHI2017}. More importantly, LIBS is significantly affected by the changes in experimental conditions (such as ablation~\cite{KEPES20201_ORTHO, KEPES_NONORTHO} and collection geometry~\cite{SHABANOV2018}). A complete list of physics-related factors responsible for the structure of spectra is beyond the scope of this work and was described elsewhere~\cite{LIBS_book_2006}. The most prominent and easily recognizable changes in the spectra’s structure can be attributed to the spectrograph and camera. Namely, spectrographs commonly differ in their spectral range and resolving power. Similarly, pixel size and related quantum efficiency of a detector affect the resolution and overall structure of detected spectra. While the detector’s response can be addressed by intensity calibration using standard light sources, corrections for the spectrograph’s impact remain elusive. 

Problems with non-trivial signal dependence on experimental conditions and instrumentation could be easily neglected for a simple qualitative analysis. Tasks like element detection and sample classification require only a wavelength position-calibrated spectrograph and a sufficient resolution for detecting desired spectroscopic lines. A high redundancy that is present in LIBS spectra~\cite{VRABEL_RBM} simplifies these tasks even more. Elements present in samples often have many characteristic lines scattered over the wavelength range. Therefore, even cost-efficient spectrographs could provide adequate performance in qualitative applications~\cite{VRABEL_SVM}. On the contrary, a reliable quantitative analysis is significantly more demanding, with the necessity of spectral intensity correction, compensation for the mentioned matrix effects, and understanding of the instrumentation's limitations. Usually, a set of known standards in the desired range is used to find the system response and to obtain a calibration curve. Such an approach has usually very limited extrapolation reliability, making it suitable only for predefined applications. An intriguing alternative to calibration curves is so-called calibration-free (CF) LIBS~\cite{Ciucci_CF}, where the plasma emission is modeled under certain conditions (local thermodynamic equilibrium, plasma stoichiometry, etc.). Despite many efforts, the CF LIBS is still not practically possible for many relevant applications and experimental conditions.

Challenges described up to this point concerned measurements on a single instrument. However, many applications would benefit from the cooperation of several systems and the combination of their outputs, i.e. the transfer of data libraries. A representative example is the ChemCam device mounted on the Curiosity Mars rover that has an exact copy operating on Earth. The compatibility of measurements from two distinct systems is the missing element for a trustworthy inter-laboratory comparison and creation of shared spectra libraries in the LIBS community. Nevertheless, several technical and physics-related limitations rule out the generally-valid transfer (as spectral ranges and resolving power of detectors may vary significantly). However, even a reasonable approximation would have great potential for practice.

A potential solution is offered by the emerging field of transfer learning. Broadly speaking, transfer learning (TL) is a collection of methods used to improve the performance of machine learning models using datasets obtained under conditions that differ from those of the target application. In the context of spectroscopy, this would entail changes in the target properties and experimental conditions. In fact, TL has been used in the context of spectroscopy to correct for changing target temperatures~\cite{yang2018libs, kaneko2021transfer, chang2020assessment}, target surface roughness~\cite{shabbir2021transfer}, and matrix~\cite{sun2021machine}. A common approach is to simply include a limited number of spectra obtained from the targets with the newly encountered properties into the training dataset~\cite{shabbir2021transfer, li2021boosting, yu2021cross, brand2021predicting}. Alternatively, a subset of the available training dataset is used, which includes the spectra only from samples that are found to be resilient to the considered changes~\cite{li2021boosting, yu2021cross, chen2019cross}. A similar approach can be applied to the features (spectral lines). That is, instead of using the whole available feature set, the features resilient to the changing conditions are selected for further analysis~\cite{yang2018libs, kaneko2021transfer, sun2021machine, brand2021predicting}. Collectively, these approaches are referred to as data-based TL.

Instead of considering the data, TL can also be performed using model-based approaches. Namely, a trained model may be fine-tuned (trained further after the initial training) on the newly obtained data, as it has been done in several spectroscopic applications~\cite{kepes2022improving, li2020multi, liu2018transfer}. Alternatively, a common representation of the old and new datasets can be found by a transformation model. This is done with the expectation that their differences are suppressed in the common representation of the two datasets. For spectroscopic applications, the construction of this common representation has been reported using transfer component analysis~\cite{pan2010domain} to address changing experimental conditions~\cite{wang2019modeling}. In an alternative approach, dimensionality reduction has been carried out separately on the two datasets, followed by adjusting for the difference in the embedding functions~\cite{chen2021calibration}.

\begin{figure}[!htb]
    \centering
    \includegraphics[width=\textwidth]{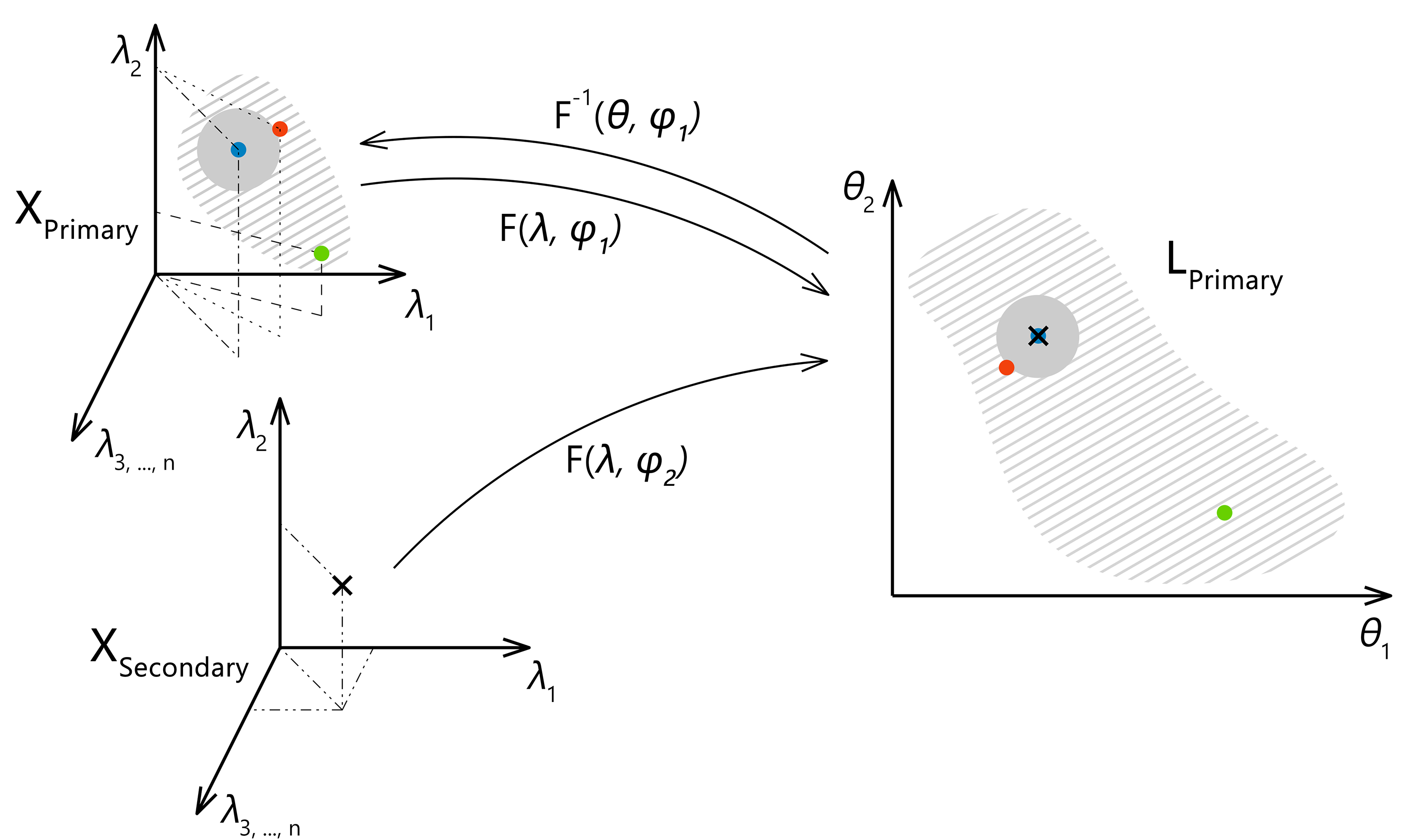}
    \caption{Schematic representation of the methodology. A latent representation $L_{Primary}$ is obtained from the Primary system and later used for mapping spectra from the Secondary system to the Primary. Points from the latent space $L_{Primary}$ can be easily reconstructed back to the space of the Primary system.}
    \label{fig:schema}
\end{figure}

In this work, we demonstrate the possibility of sharing a spectra library between the original Primary system and (possibly many) Secondary systems. For the transfer, we use a pipeline that consists of a variational autoencoder (VAE), used for obtaining a latent representation of the Primary system spectra and an artificial neural network (ANN) for mapping corresponding spectra from the Secondary system to this latent representation (see Figure \ref{fig:schema}). We demonstrate that this methodology is able to transfer spectra between two systems even when the Secondary system covers a narrower spectral region (given by a sufficient amount of training data). Overall, the approach is expected to be generalizable to out-of-sample cases where the target domain is beyond shared calibration samples or the simultaneous measurement regime. The envisioned application would connect many laboratories to the Primary system and enable better inter-laboratory cooperation and comparison.
A superficially similar problem has already been addressed~\cite{chen2016calibration}; however, there are significant differences in both the approach and the data. The mentioned work considers only spectrometers measuring on the same wavelength ranges, while we allow for different resolutions and spectral ranges (see Figure \ref{fig:means}). This poses a much more complex problem since it requires the model to capture nontrivial relations between different segments of the measured range. Additionally, the term autoencoder, used in the name of the method presented in~\cite{chen2016calibration}, is slightly confusing as it refers only to the similarity in the model architecture with autoencoder architecture (i.e. presence of the bottleneck). However, it was not trained or used as the autoencoder (i.e. for the dimension reduction and reconstruction) but as a standard ANN.

\section{Experiment and Data}
\label{sec:Data}
\noindent
The instrumental setup consisted of two spectrometers (each with its own collection system) synchronized for simultaneous measurements and shared the rest of the instrumental setup. Both spectrometers were Czerny-Turner type, with partially overlapping spectral ranges of different resolutions and sizes. Henceforth, we refer to the spectrometer with a broader spectral range, its collecting optics, and the rest of the instrumentation (laser, delay generator, …) as an experimental Primary system. Analogically, the Secondary system is the set containing a Czerny-Turner spectrometer with a narrower spectral range and shared parts. The collected datasets will be denoted by $X_{Primary}$ and $X_{Secondary}$, respectively.

To ensure a large variability of training data, we selected a heterogeneous rock sample (rare-element granitic pegmatite, further described below) and measured it in a mapping regime (i.e. raster on the sample surface was formed and one spectrum was measured from each spot).  Each row (spectrum) $x_{i, Primary}$ corresponds to the row $x_{i, Secondary}$, i.e., there is a one-to-one correspondence between the measurements, which were obtained from the same laser-induced plasma.

The test rock sample was cut from a strongly heterogeneous rock, classified as granitic pegmatite (locality Maršíkov D6e). Pegmatites are vein rocks specific by an extreme degree of chemical and textural fractionation, resulting in large contents of chemical elements that are otherwise rare in Earth's Continental Crust. These include especially e.g. Li, Be, F, Rb, Cs, Ta, and Nb. The studied rock sample is mineralogically dominated by three minerals: quartz, albite, and muscovite. In subordinate amounts, it contains Fe-Mn garnet (almandine-spessartine), beryl, and Nb,Ta-oxides (columbite, fermite, microlite).

\begin{figure}[!htb]
    \centering
    \includegraphics[width=0.8\textwidth]{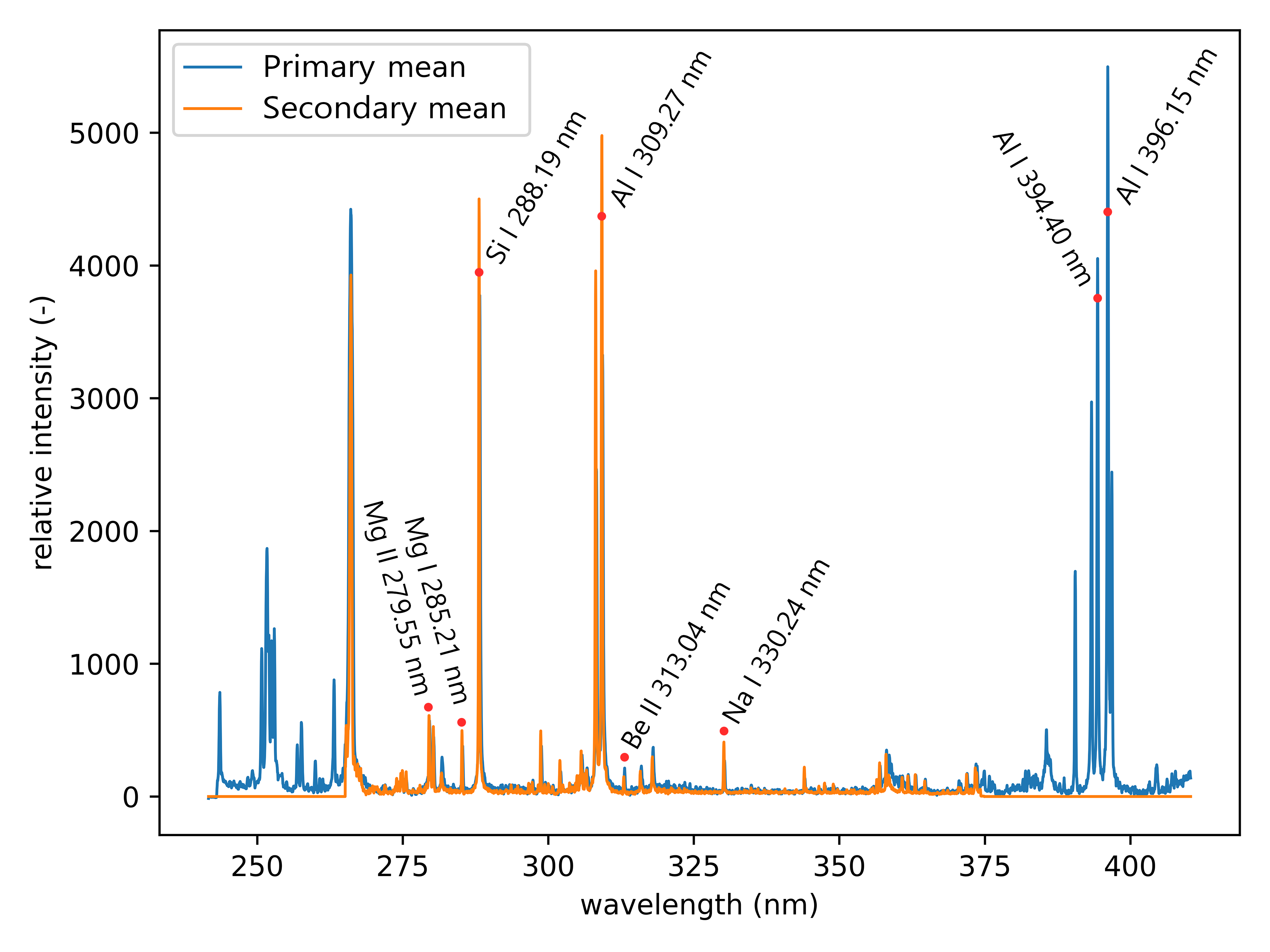}
    \caption{Average spectra of the Primary and Secondary system from the test dataset. Significant lines that are relevant to the sample composition are labeled.}
    \label{fig:means}
\end{figure}

All LIBS measurements were performed on the Firefly LIBS system (Lightigo, Czech Republic). The Firefly was equipped with a diode-pumped solid-state laser (266 nm, 5 ns, 50 Hz, 5 mJ) used for sample ablation and plasma generation. The laser beam was focused onto the sample (30 \textmu m spot size) and plasma emission was then collected by a wide-angle objective and through an optical fiber bundle (400 \textmu m core diameter), where each of the fibers collected radiation from a slightly different part of the plasma. Plasma emission was transferred to both Czerny-Turner spectrometers. The spectra range of the  Primary system ranged from 245 to 407 nm with resolution varying from 0.035 to 0.046 nm and for the Secondary secondary system from 260 to 370 nm with resolution varying from 0.023 to 0.032 nm (see Figure \ref{fig:means}). The slit size was 25 µm for both spectrometers. Samples were placed onto a computer-controlled motorized stage. Spectra were acquired for each sampling position resulting in a raster with 40 µm spacing for training and 20 µm for the test dataset, in both X and Y axes.

The collected data from both systems were separated into three datasets - the training dataset (used for model training), the validation dataset (used for hyperparameter optimization and model validation during the training process), and the test dataset (used for final one-shot testing). To ensure that the results are representative of the performance on unseen data, each dataset corresponds to a separate measurement (out-of-sample evaluation) of a hyperspectral image. Their respective dimensions are $560 \times 560$ (training), $266 \times 500$ (validation), and $500 \times 500$ (test). Note, that the test dataset was measured on a different day than the training and validation datasets. The sample photo with highlighted locations for each of the datasets is available in the supplementary materials. 

The training/validation/test datasets consist of 313,600/133,000/250,000 spectra (about 45-20-35\% ratio) with 4062 and 3872 features for Primary and Secondary systems, respectively. For example, this means that the dimension of the Primary training dataset is $X_{Primary} \in \mathbb{R}^{313600 \times 4062}$ and $X_{Secondary} \in \mathbb{R}^{313600 \times 3872}$ for Secondary. The combined size of all the datasets is roughly $22$ GB.

\begin{figure}[!htb]
    \centering
    \includegraphics[width=0.8\textwidth]{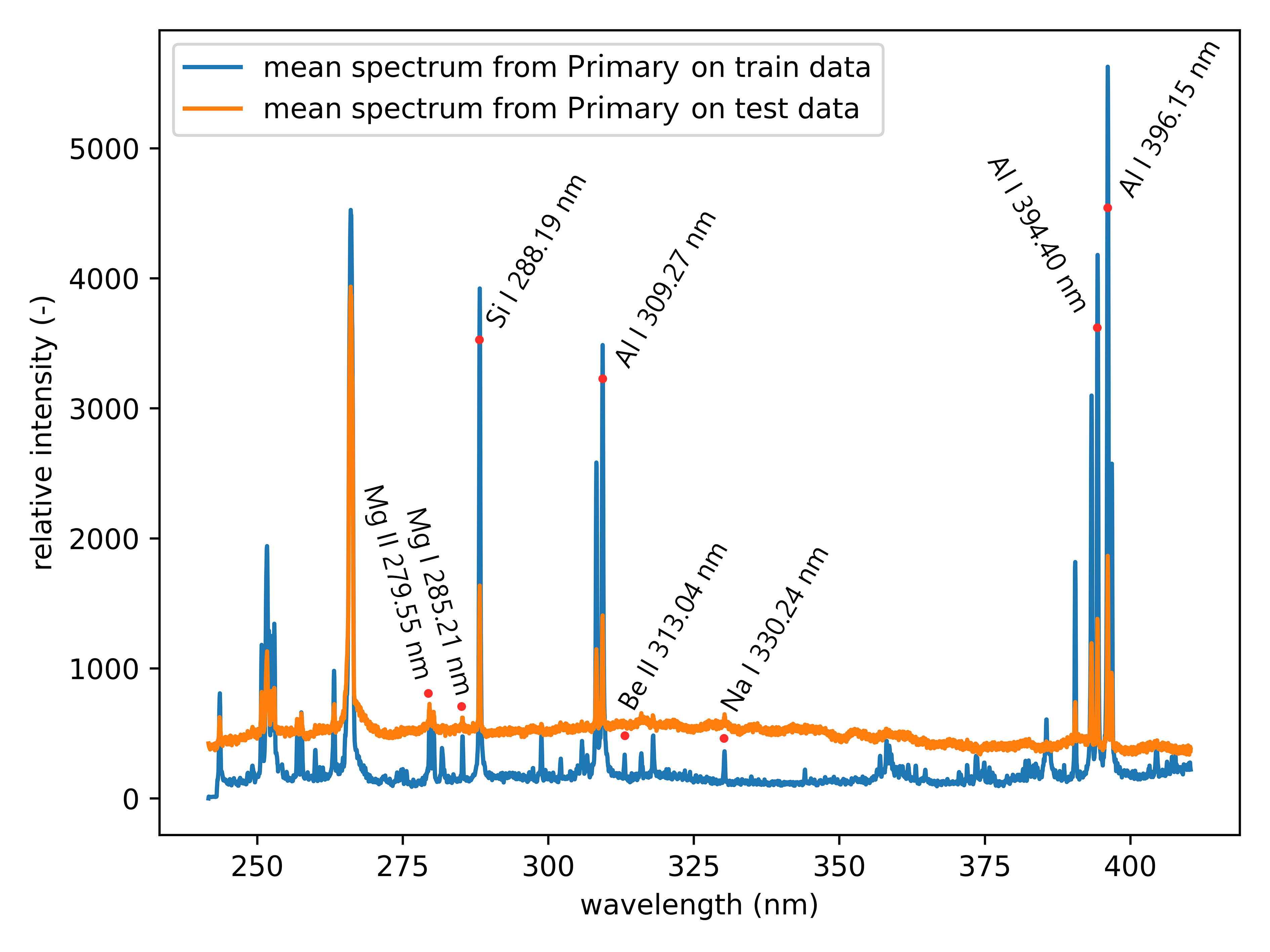}
    \caption{Average spectra from the Primary system before preprocessing. Comparison of test and training datasets. There is a notable shift in the baseline.}
    \label{fig:shift}
\end{figure}

All spectra were baseline corrected by performing a sliding minimum operation followed by a sliding Gaussian mean smooth (window size $100$ and smooth parameter $50$)~\cite{kepevs2018influence}. In addition, each dataset was individually mean-centered and scaled to feature-wise unit variance (standard scaling). This was done to compensate for the significant shift in the baseline of the spectra from the training and the test datasets (see Figure \ref{fig:shift}). Since the dataset's original mean and variance can be saved, this transformation is lossless. All results are presented in the original form.

\section{Methodology}
\label{sec:Met}
\noindent
The envisioned application of the proposed methodology is to create a Primary system that will be potentially shared by multiple Secondary systems. Considering the non-matching spectral ranges and resolutions of two distinct experimental systems, the corresponding spectra $X_{Primary}$ and $X_{Secondary}$ will differ significantly. Thus, to transfer spectra between systems, we aim to find a mapping $f: X_{Secondary} \rightarrow X_{Primary}$ that gives a corresponding Primary spectrum for each measurement in the Secondary system. Since the mapping between the two spaces of a different dimension and spectral range coverage cannot be generally found, we obtain an effective mapping $f^{\prime}: X_{Secondary} \rightarrow X_{Primary^{\prime}}$ and utilize it for practical applications.

\begin{figure}[!htb]
    \centering
    \includegraphics[height=0.5\textheight]{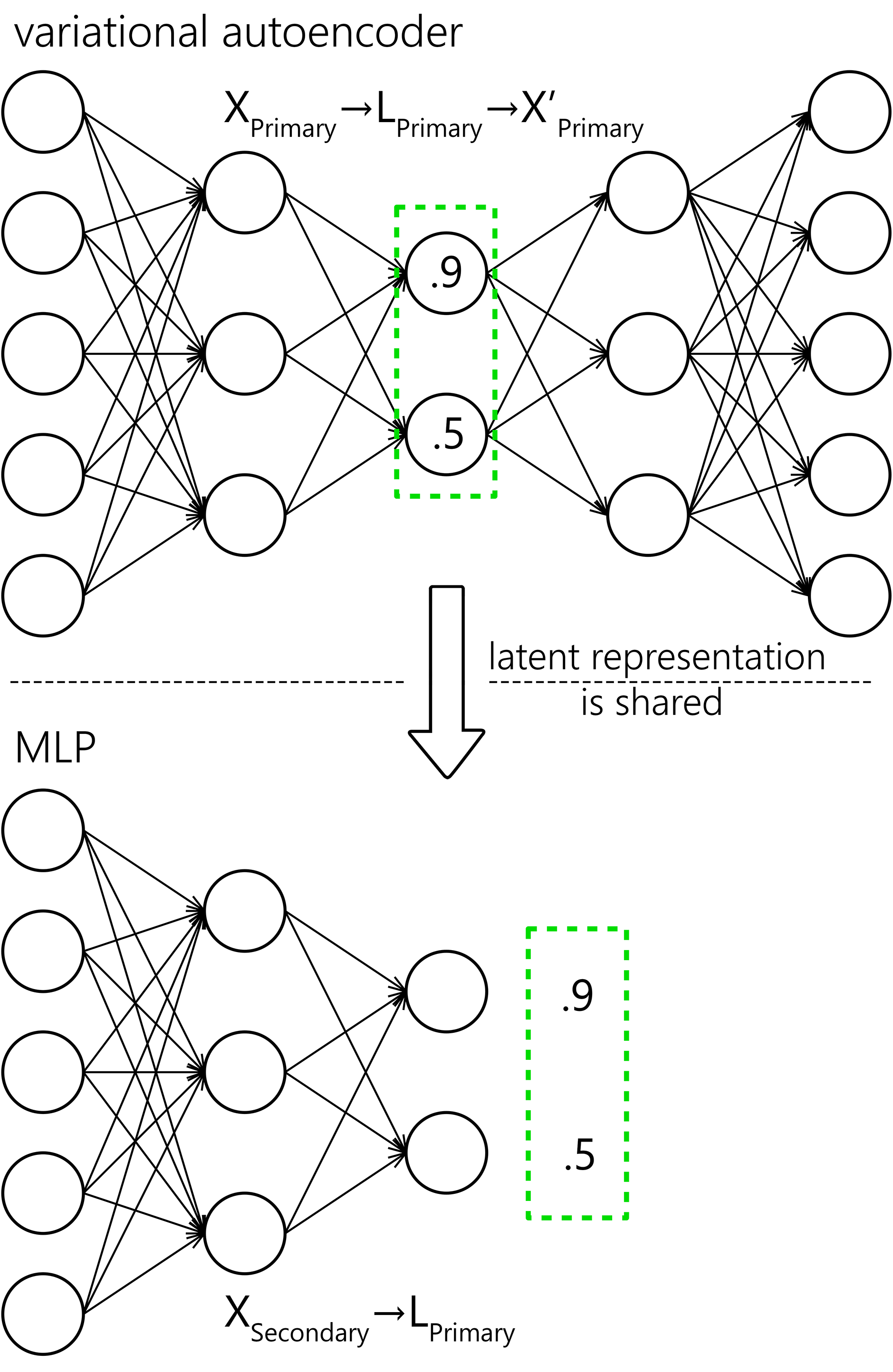}
    \caption{Schematic representation of both mappings. The latent encoding of the $X_{Primary}$ is shared with MLP as ground truth for the training. Values in the latent representation and neuron counts are only illustrational.}
    \label{fig:schema2}
\end{figure}

We propose a two-step approach: using a VAE (see section \ref{sec:Model}), we obtain a (low-dimensional) latent representation of the data (denoted as $L_{Primary}$). This representation, along with the decoder, is shared and stored centrally. In the second step, repeated for every Secondary system, we train a multilayer perceptron (MLP) to map the $X_{Secondary}$ spectra to the latent space $L_{Primary}$ (as depicted in Figure \ref{fig:schema2}). Finally, by combining the shared decoder with the newly trained MLP we have an approximation of the desired mapping $f$, $f^{\prime}: X_{Secondary} \rightarrow X_{Primary}^{\prime}$. Note that for the proposed methodology to work, the one-to-one correspondence of training measurements on the Primary and the Secondary system is required. This is most easily achieved by obtaining the two datasets concurrently from the same plasma (as was done in this work). Nevertheless, the alignment could be achieved even for more general cases, e.g., via optimal transport. This option will be explored in our future work.

\subsection{Evaluation}
\label{sec:Eval}
\noindent
We evaluate the performance (i.e. the discrepancy between the Primary spectra and predicted counterparts from the Secondary system) of the methodology from the two basic points of view that are relevant to the LIBS applications; first the quantitative, and second, the qualitative analysis. For the purposes of estimating the negative impact of the differences between $X_{Primary}$ and predicted  $X_{Primary}^{\prime}$ we introduce two measures of spectra distance:

Euclidean distance:
\begin{equation}
    Euclid = \sqrt{\sum_{i=1}^{M}(y_{i} - \hat{y}_{i})^2},
\end{equation}
where $M$ is the number of measured wavelengths, $y_{i}$ is the actual value on the $i$-th wavelength and $\hat{y}_{i}$ is the predicted value.
Cosine distance:
\begin{equation}
    cosine = 1 - \frac{Y \cdot \hat{Y}}{\lvert Y \rvert \cdot \lvert \hat{Y} \rvert},
\end{equation}
where $Y$ is the original spectra interpreted as a vector and $\hat{Y}$ is the predicted one. It ranges from $1$ to $-1$, where values closer to $-1$ indicate greater similarity, $0$ indicates orthogonality and $1$ indicates opposite values. We then average the results gained for each spectra pair to obtain a single value that represents results on the entire dataset.

Additionally, we also calculate the relative squared error ($RSE$) as:
\begin{equation}
    RSE = \frac{\sum_{i=1,j=1}^{M,N} (y_{i,j} - \hat{y}_{i,j})^2}{\sum_{i=1,j=1}^{M,N} (\overline{y}_{i} - \hat{y}_{i,j})^2},
\end{equation}
where $M$ is the number of measured wavelengths, $N$ is the number of measured spectra, $y_{i, j}$ is the actual value on the $i$-th wavelength of the $j$-th spectra, $\hat{y}_{i, j}$ is the predicted value and $\overline{y}_{i}$ is the mean of the actual values on the $i$-th wavelength. Intuitively, $RSE$ represents the performance of the model relative to the performance of a na\"ive-baseline prediction. Here, we chose the mean spectrum of the test dataset as the na\"ive-baseline prediction. The RSE ranges from $0$ to $\inf$, where $0$ indicates a perfect reconstruction, $1$ indicates the same performance as the na\"ive-baseline, and values greater than $1$ indicate worse performance. 

Lastly, for the impact on the qualitative analysis, we train a k-means clustering algorithm on the $X_{Primary}$ train dataset and compare its predictions on the $X_{Primary}$ test and the corresponding $X_{Primary}^{\prime}$ test measurements (i.e. spectra transferred from the Secondary to the Primary system). A simple accuracy metric was used:
\begin{equation}
    k-score = \frac{\sum_{j=1}^{N} eq(\theta(y_{j}), \theta(\hat{y}_{j}))}{N},
\end{equation}
where $N$ is the number of measured spectra, $\theta(y_{j})$ and $\theta(\hat{y}_{j})$ represent the predicted label on the $j$-th spectra from the $X_{Primary}$ and $X_{Primary}^{\prime}$ dataset respectively, and eq: $\mathbb{N} \rightarrow \mathbb{N}$:
\begin{equation}
eq(x, y)=
   \begin{cases}
       1 & \text{if } x = y, \\
       0 & \text{otherwise.}
   \end{cases}
\end{equation}

We benchmark the model against common multivariate regression baselines - a na\"ive regressor (denoted as mean baseline), predicting the mean spectra from the training dataset and k-nearest neighbors regressor (denoted as KNN baseline), described in more detail in the following section. Additionally, we also compare to a MLP directly trained to predict the $X_{Primary}$ spectra from the $X_{Secondary}$ (denoted as MLP baseline).

\subsection{Models}
\label{sec:Model}

\subsubsection{Multilayer perceptron (MLP)}
\label{sec:mlp}
\noindent
ANNs are computational models defined by their architecture and learning~\cite{lecun2015deep}. In general, they are composed of formal neurons (defined by a set of weights and an activation function) arranged into interconnected layers of set sizes - one input, one output, and possibly other hidden. These connections form the architecture of the network. ANNs can approximate any continuous function~\cite{cybenko1989approximation} that relates the output of the network to its input. The output of the network is computed by a subsequent application of layers. Weights (i.e. parameters) of the network have to be learned in order to make relevant predictions, which is usually done by a stochastic gradient descent algorithm with backpropagation~\cite{lecun1988theoretical}.
MLP~\cite{rumelhart1985learning} is a well-established~\cite{gardner1998artificial, kotsiopoulos2021machine} type of neural network architecture that is feedforward (i.e. connections between the neurons do not form cycles) and fully-connected (each neuron within every layer is connected to all neurons in the subsequent layer, with the exception in the output layer).
In this work, the model parameters (architecture and learning parameters) were optimized on the validation data with a hyperband algorithm~\cite{omalley2019kerastuner}. The considered domains, as well as other fixed parameters, were selected heuristically, using prior experience with processing of spectroscopic data, along with non-exhaustive manual experimentation. The MLP is composed of two layers. For the first (second) layer 2048, 1024, 512, 128 (1024, 512, 128) neurons were considered, 128 (512) was chosen by the optimization algorithm as best. The leaky ReLU activation function was used in every layer except the last one which utilized a linear activation. L2 regularization was used in every hidden layer in order to increase the generalization capabilities of the model. For training, the Adam optimizer was selected. The learning rate was optimized to 1e-4 from the following options: 1e-3, 1e-4, 1e-5. The model was trained for 100 epochs with early stopping set on validation loss and batch size of 128. To reiterate, the inputs of the model are the $X_{Secondary}$ spectra and the outputs are mapped as closely as possible to the corresponding $L_{Primary}$ embeddings.
The MLP baseline model was fully determined by the parameters of the MLP and the autoencoder. The $X_{Secondary}$ spectra are the input and the desired outputs are the $X_{Primary}$ spectra directly.

\subsubsection{Autoencoder (AE)}
\label{sec:ae}
\noindent
Autoencoders~\cite{kramer1991nonlinear} are unsupervised deep learning models commonly used for tasks such as dimensionality reduction~\cite{wang2014generalized}, data denoising~\cite{vincent2008extracting}, and others~\cite{bank2020autoencoders}. Autoencoders are trained to encode and reconstruct the input dataset with the stipulation that at some point in the topology of the network there is a bottleneck - a layer with a limited number of neurons. Using the bottleneck, we can separate the network into two parts - the encoder (which can be interpreted as a MLP creating the $L_{Primary}$ encodings) that consists of the bottleneck and the layers preceding it, and the decoder (a MLP reconstructing $X_{Primary}$ from ${L}_{Primary}$), made up of the layers after the bottleneck. The bottleneck is where we can extract a low-dimensional latent representation of the original input (encoding). It should be noted that the representation obtained from the autoencoder is lossy. For the proposed methodology we used a Variational Autoencoder (VAE)~\cite{VAE_welling} that differs from the vanilla AE by a regularization of the bottleneck in order to achieve desirable characteristics of the latent space to allow the generation of new data, as well as help with generalization. The aforementioned characteristics are continuity (two “close” points should give similar results) and completeness (for a previously chosen distribution, any point sampled from it should give meaningful results). In practice, VAE is trained to predict a distribution (defined by its expected value and its variance), rather than predicting the latent encoding directly. We then also regularize the distribution in the latent space to fit the normal distribution centered at zero with unit variance using Kullback–Leibler divergence~\cite{kingma2013auto}.
Similarly to the MLP, the model parameters were optimized on the validation data with a previously mentioned hyperband algorithm. The autoencoder is composed of five layers, mirrored around the bottleneck. For the first and last layer we considered the following options: 2048, 1024, 512, and 128, out of these 1024 was chosen by the optimization algorithm as best. For the second and fourth (second to last) layers we considered 1024, 512, and 64 neurons. The best option was 512. The bottleneck was optimized to 64 neurons out of 3, 8, 32, 64. The leaky ReLU activation function was used in every layer except the last and the bottleneck, which utilized a linear activation. L2 regularization was used in every hidden layer (except the bottleneck) in order to increase the generalization capabilities of the model. The Kullback–Leibler divergence was scaled by 4-cycle linear annealing~\cite{fu2019cyclical} going from 0 to 0.5. For training, the Adam optimizer was selected. The learning rate was optimized to 1e-4 from the following options: 1e-3, 1e-4, 1e-5. The model was trained for 50 epochs with a batch size of 128. Both the input and the expected output of the model is $X_{Primary}$.

\subsubsection{K-Nearest Neighbors Regressor (KNN)}
\label{sec:knn}
\noindent
The KNN algorithm~\cite{altman1992introduction} is a simple non-parametric regression and classification algorithm. To make predictions, it memorizes the entire training dataset (both the inputs and the desired outputs), and uses a distance metric (Euclidean distance in our case) to find the $k$ (10 in our case, which was found by an optimization from values 2, 3, 4, 5, 10, 15, 20) most similar training samples. The prediction is gained either by a majority vote for classification or as a mean of the memorized outputs for regression. We optimized the distance metric and the number of neighbors to minimize the average Euclidean distance between spectra from the validation dataset.

\subsubsection{K-Means}
\label{sec:kmeans}
\noindent
K-means (na\"ive k-means, Lloyd’s algorithm)~\cite{1056489} is an unsupervised clustering algorithm designed to partition the data into $k$ distinct groups. Membership of a sample to a cluster is decided based on the (Euclidean) distance from iteratively refined cluster centers.
The number of partitions was optimized on the test dataset to minimize the silhouette score~\cite{rousseeuw1987silhouettes}, a metric based on the inter-cluster distance (average distance between the samples within the same cluster) and the distance of all samples to the next nearest cluster. Due to the computational cost, only a randomly selected subset of 25000 (10\%) was used. $K = 4$ was selected as the optimal number from the considered options: 3, 4, 5, 6, 7, 8, 9, 10, 15, 20.

\section{Results and Discussion}
\label{sec:Results}
\begin{figure}[!htb]
    \centering
    \includegraphics[width=\textwidth]{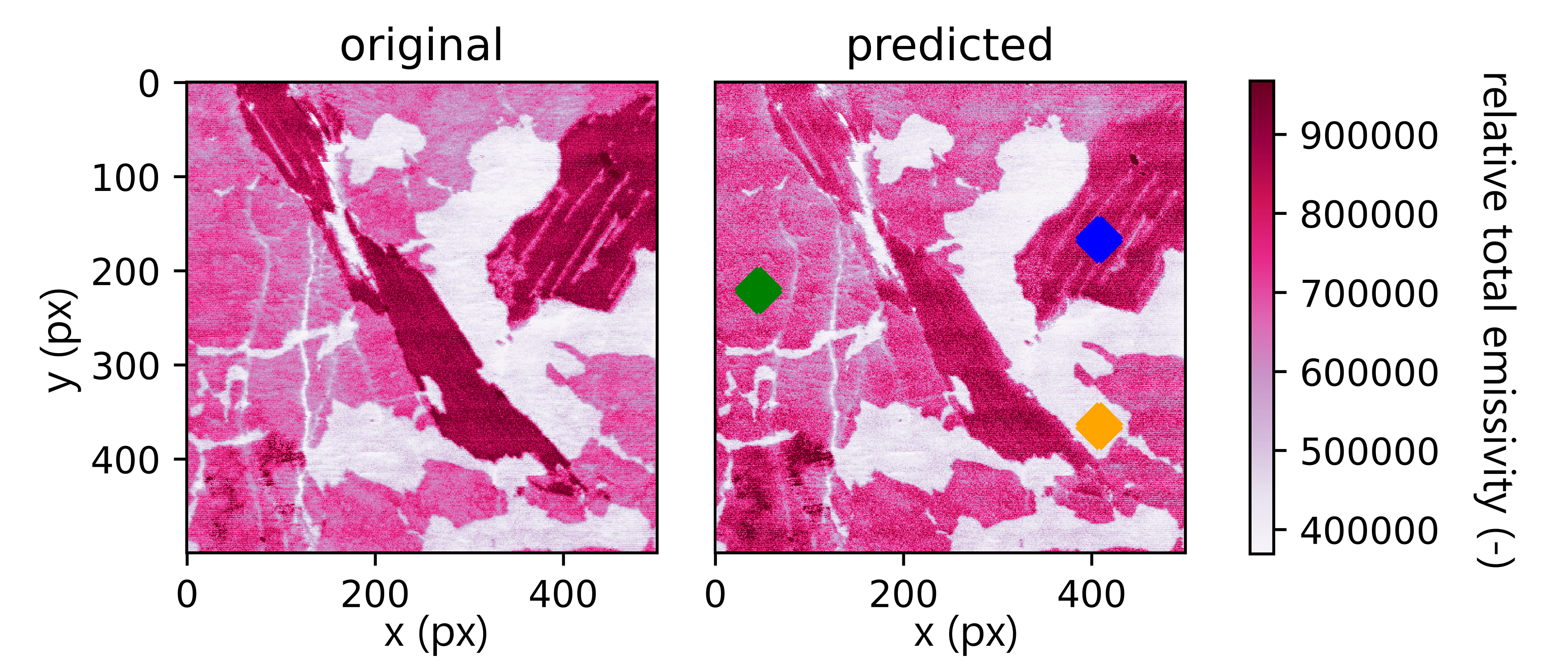}
    \caption{Comparison of total emissivity maps of $X_{Primary}$ (original) and $X_{Primary}^{\prime}$ (predicted). There is a visible decrease in overall intensity. One representative for each visible spectra category is marked. First and last quantiles were aggregated together to increase visibility. Three color-marked spots are used for representative spectra comparison below.}
    \label{fig:emissivity}
\end{figure}

\noindent
In this section, we present the performance of the proposed method along with the baseline models in a one-shot evaluation on the test dataset. Results of the evaluation metrics (as described in the section \ref{sec:Eval}) are listed in Table~\ref{table:errors}. Both ANN-based models significantly overperformed other considered models, in all the error metrics, as well as in train and prediction times (see Table~\ref{table:times}). While the MLP baseline model has slightly higher scores, it requires significantly more data to be shared and stored between the participating systems (approximately 50 times more). Moreover, the selected approach (VAE+MLP) possesses a generative property that is preferred for a future extensivity and generalizability of the proposed transfer methodology.
In the following paragraph, we study strengths and weaknesses of the methodology. The total emissivity map of the $X_{Primary}$ dataset compared to the $X_{Primary}^{\prime}$ can be seen in Figure \ref{fig:emissivity}. It should be noted that the spectra with intensity in the first or last quantiles are displayed with the same color (aggregated). This is done to keep the clarity of the sample structures even with outliers present in the data (the same approach is repeated in each subsequent hyperspectral image).

\begin{table}[htb!]
\centering
\begin{tabular}{|c| c c c c|} 
 \hline
  X & Euclid & cosine & RSE & k-score \\ [0.5ex] 
 \hline
 VAE + MLP & 4658.72 & -0.9919 & 0.1219 & 97.578 \\ 
 MLP baseline & 4201.44 & -0.9931 & 0.1021 & 97.9768 \\
 KNN baseline & 31444.34 & -0.5634 & 4.2585 & 32.2456 \\
 Mean baseline & 22188.85 & -0.7918 & 2.3592 & 27.1064 \\ [1ex] 
 \hline
\end{tabular}
\caption{Comparison of the methodology performance with selected baselines.}
\label{table:errors}
\end{table}

\begin{table}[htb!]
\centering
\begin{tabular}{|c| c c|} 
 \hline
 X & approx. training time\tablefootnote{GPU was used when appropriate} & approx. prediction time \\ [0.5ex] 
 \hline
 VAE + MLP\tablefootnote{training time of the autoencoder is not considered} & 80 s & 15 s \\ 
 MLP baseline & 300 s & 15 s \\
 KNN baseline & 3 s & 3600 s \\
 Mean baseline & 3 s & 3 s \\ [1ex] 
 \hline
\end{tabular}
\caption{Comparison of the methodology training times with selected baselines.}
\label{table:times}
\end{table}

\begin{figure}[!htb]
    \centering
    \includegraphics[height=0.5\textheight]{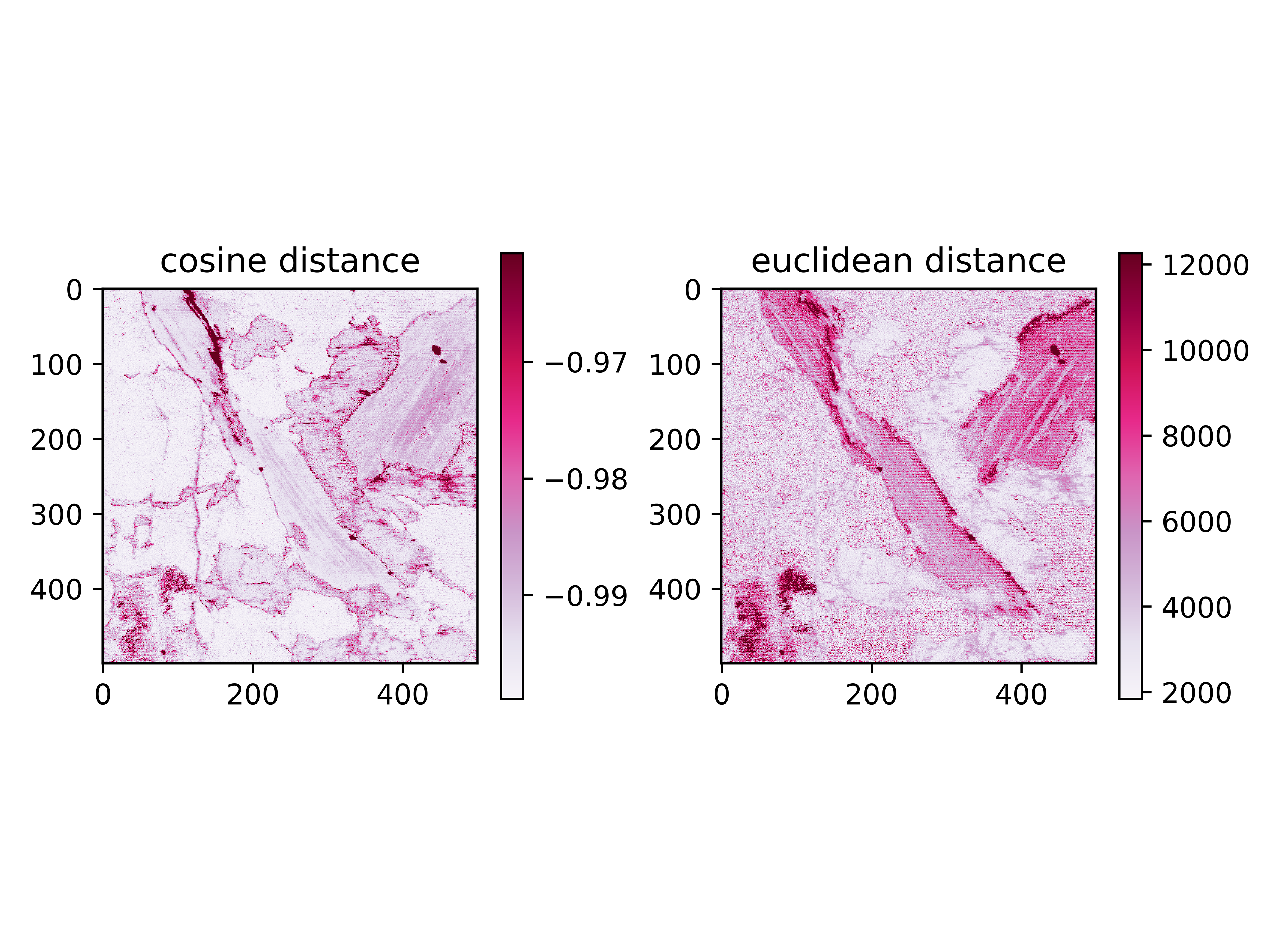}
    \caption{Spatial distribution of the euclidean and cosine distances. First and last quantiles were aggregated together to enhance the contrast.}
    \label{fig:error_maps}
\end{figure}

The spatial distribution of the transfer error can be seen in Figure \ref{fig:error_maps}. Ideally, the error would be spatially invariant; however, it is clear that this is not the case. Spectra with the highest error are assembled on the borders of distinct matrices. This follows from the underrepresentation of boundary spectra in the model, and contamination of the emission signal by the previous measurements.

\begin{figure}[!htb]
    \centering
    \includegraphics[width=0.8\textwidth]{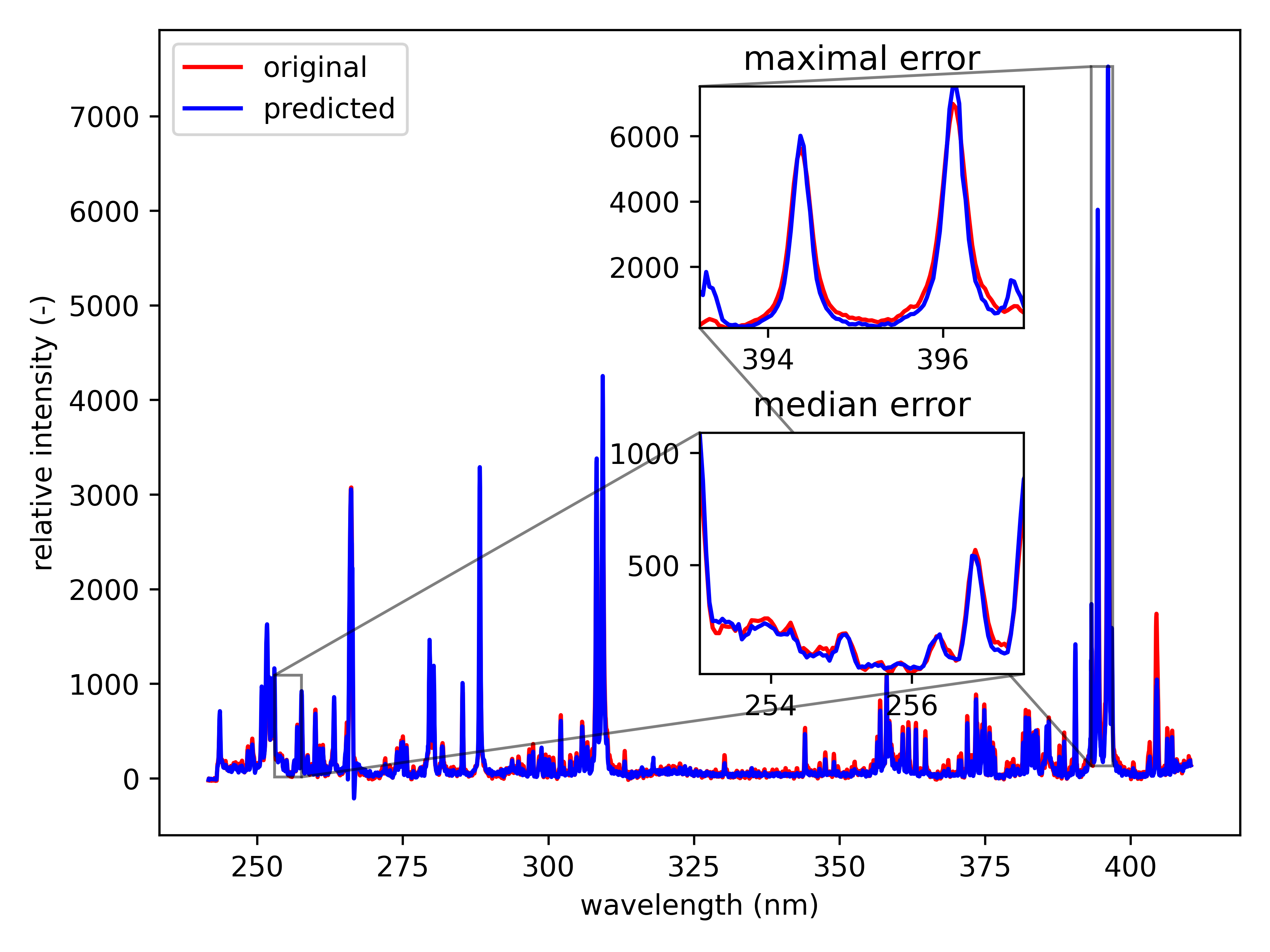}
    \caption{Blue spot,  from Fig. 5 (coordinates 408, 167). Muscovite (ideally KAl3Si3O10(OH)2; composition:: K 9.8 wt.\%, Al 20.3 wt.\%, Si 21.2 wt.\%, O 48.2 wt.\%, H 0.5 wt.\%). Original spectrum from the Primary system and prediction from the Secondary system.}
    \label{fig:blue}
\end{figure}

\begin{figure}[!htb]
    \centering
    \includegraphics[width=0.8\textwidth]{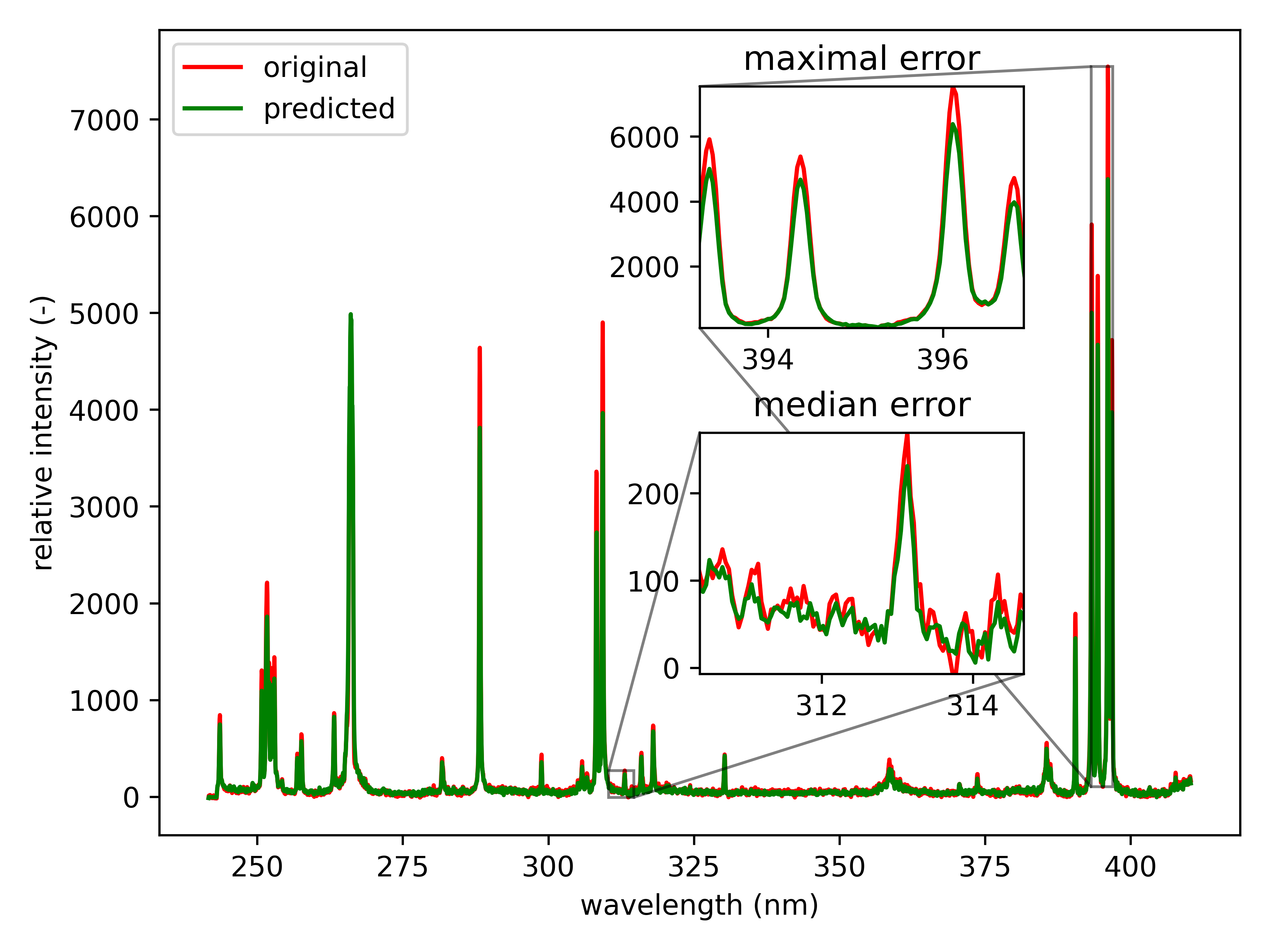}
    \caption{Green spot from Fig 5. (coordinates 46, 221). Albite (ideally NaAlSi3O8; element contents: Na 8.8 wt.\%, Al 10.3 wt.\%, Si 32.1 wt.\%, O 48.8 wt.\%). Original spectrum from the Primary system and prediction from the Secondary system.}
    \label{fig:green}
\end{figure}

\begin{figure}[!htb]
    \centering
    \includegraphics[width=0.8\textwidth]{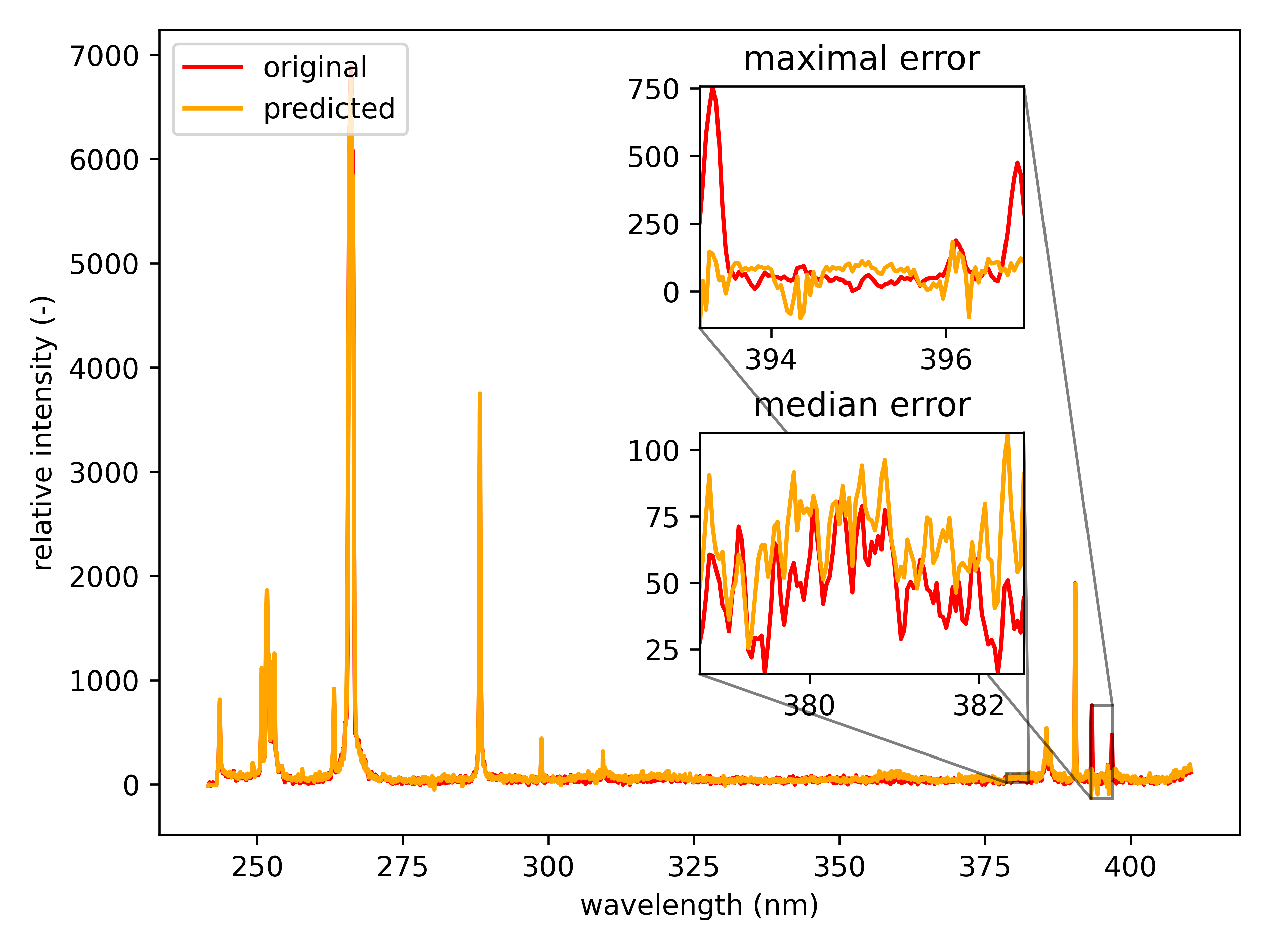}
    \caption{Orange spot from Fig 5. (coordinates 408, 365). Quartz (ideally SiO2; element contents: Si 46.7 wt.\%, O 53.3 wt.\%). Original spectrum from the Primary system and prediction from the Secondary system.}
    \label{fig:orange}
\end{figure}

Qualitatively, the transfer was highly successful, the sample topology is well-preserved in the predicted map, see Figure \ref{fig:emissivity}b. However, some important parts (spectral lines) are predicted imperfectly. For the applications related to qualitative analysis, these errors are not significant, but for quantitative analysis this could be considerable. To further investigate the prediction error and its wavelength dependency, we selected three representative spots, each from a different matrix present in the sample, see Figure \ref{fig:emissivity}, and compared original and predicted spectra from these spots, see Figures \ref{fig:blue}, \ref{fig:green} and \ref{fig:green}. Predictions of representative spectra depicted in Figures \ref{fig:blue} and \ref{fig:green} show a good performance of the transfer methodology, where only a minor-to-moderate discrepancy between specific line intensities is present. However, in Figure \ref{fig:orange}, a significant reconstruction error can be seen between 390 nm and 410 nm, where the spectral peaks are predicted as negative. This error is representative for most of the spectra of the same matrix. It is likely a consequence of the significant increase in the background signal of the corresponding test spectra for this matrix, since the aforementioned peaks completely blend with the background noise in the test dataset (see Figure \ref{fig:shift}).

\begin{figure}[!htb]
    \centering
    \includegraphics[width=0.8\textwidth]{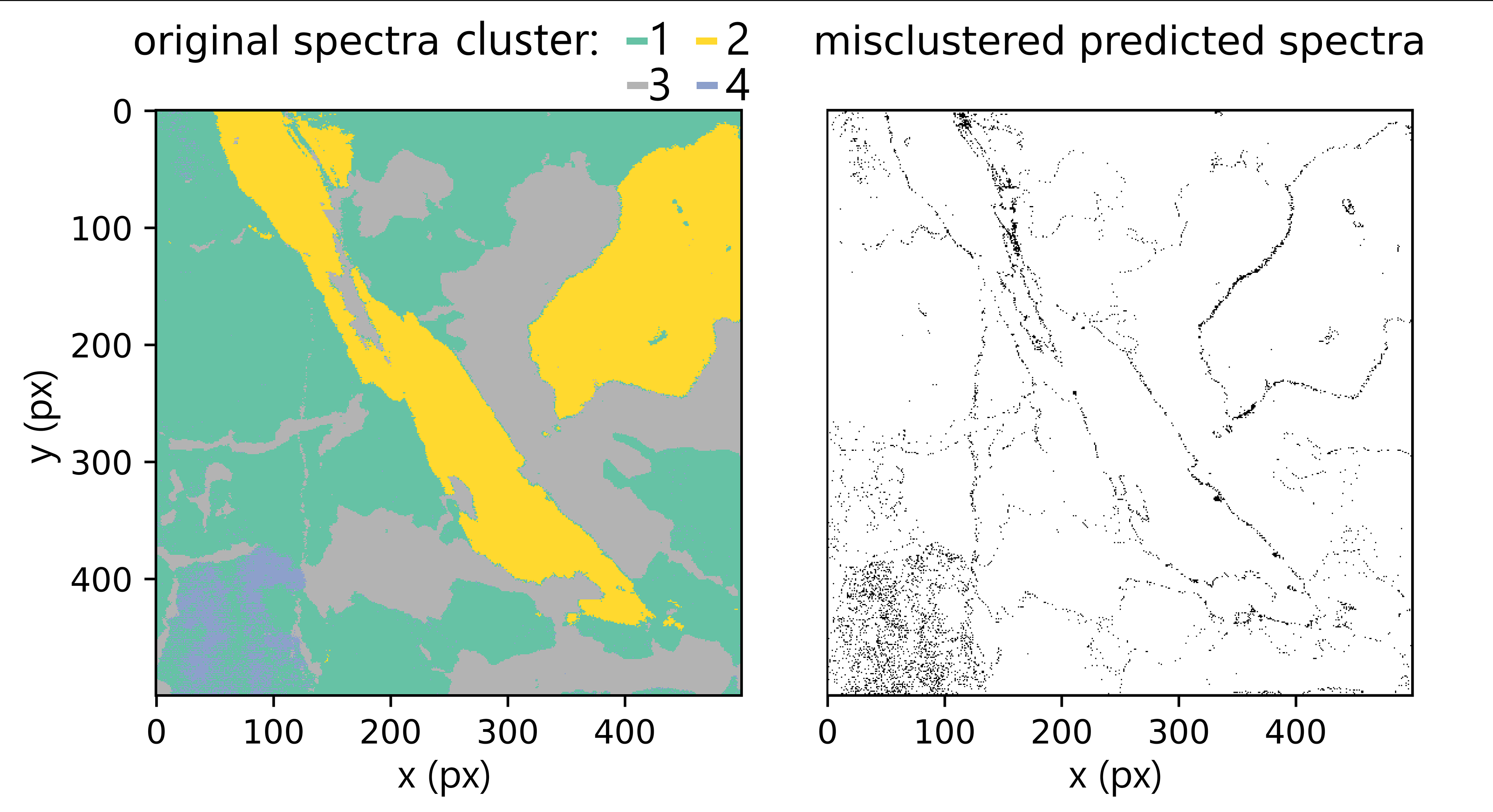}
    \caption{Labels of the $X_{Primary}$ (original) dataset as predicted by k-means (k = 4), along with the highlighted differences to the $X_{Primary^{\prime}}$ (predicted) dataset.}
    \label{fig:kmeans}
\end{figure}

Lastly, we present the results from the k-means experiment described in the methodology section. The hyperspectral images of the test dataset were clustered into four clusters given by the k-means model built on the training dataset. We compared predictions on the Primary test and Secondary test images and plotted misclustered spots in Figure \ref{fig:kmeans}. The k-score was 97.578, which demonstrates a great potential for qualitative applications of the methodology (including classification tasks). Furthermore, results show the robustness of the method and the ability to generalize when faced with previously unseen data (even if they are coming from a different measurement and samples with different matrix/composition ratios). However, they also reaffirm the suspected limitation of predicting the spectra on the borders of distinct matrices, which follows from the underrepresentation of corresponding spectra in the model.

\section{Conclusion}
\label{sec:Concl}
\noindent
We have demonstrated the possibility of transferring spectra between two distinct LIBS systems in a simplified setup, where both systems measure shared standards simultaneously. This approach may serve as an initial point for a more general transfer between two systems with lower demands on shared calibration standards. The methodology consists of a two-step computation, where first, a latent representation of the Primary system spectra is found (using variational autoencoder), and second, spectra from the Secondary system are mapped to this latent space. This procedure was trained and validated on large hyperspectral maps of a heterogeneous rock sample. It was demonstrated that such a transfer is possible and performs well even on unseen data from a different measurement. The performance and limitations of the transfer were evaluated by studying several metrics relevant to qualitative and quantitative LIBS analysis. We found that the spectra transfer will have only a negligible effect on the clustering. Considering the quantitative analysis, the transfer error is spatially dependent and may negatively affect quantitative predictions about the sample composition in the underrepresented regions (mostly boundaries of matrices in the sample).

\section{Acknowledgement}
\label{sec:Ack}
\noindent
JV is grateful for the support provided by the grant CEITEC-K-21-6978 and CEITEC VUT-J-22-8099 from the Brno University of Technology. Grant CEITEC-K-21-6978 is realised within the project Quality Internal Grants of BUT (KInG BUT), Reg. No. CZ.02.2.69 / 0.0 / 0.0 / 19\_073 / 0016948, which is financed from the OP RDE. JK gratefully acknowledges the support of the grant FSI-S-20-6353 from the Faculty of Mechanical Engineering, Brno University of Technology.

\bibliographystyle{elsarticle-num} 
\bibliography{cas-refs}

\appendix
\newpage

\section{Data and Code}
\label{sec:appendix}
\noindent
The used datasets are available at \url{http://dx.doi.org/10.6084/m9.figshare.20713504}.
The code is available at \url{https://github.com/LIBS-ML-team/libs-transfer-library}.

\section{Additional Figures}

\begin{figure}[!htb]
    \centering
    \includegraphics[width=0.8\textwidth]{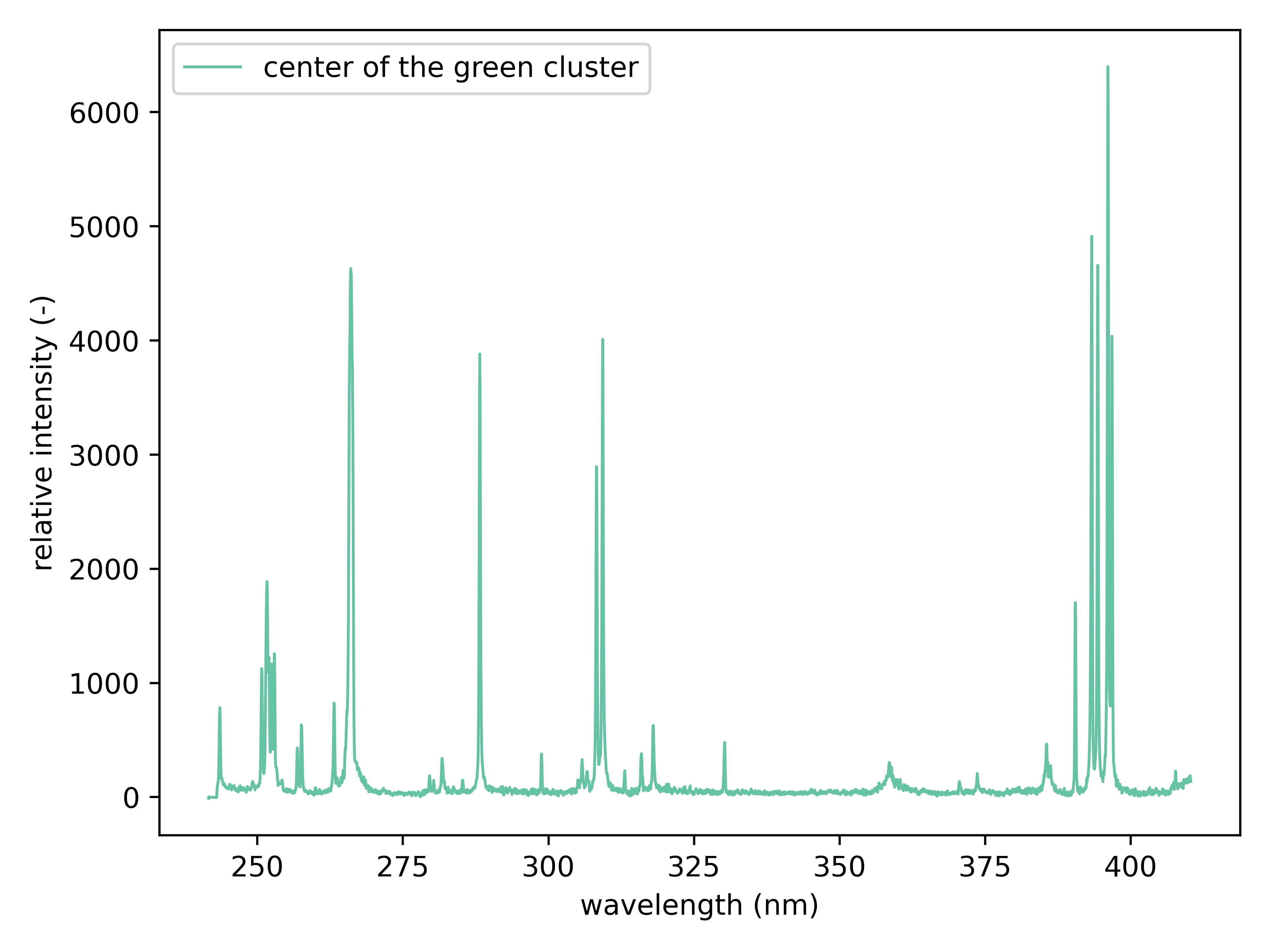}
    \caption{Center of the green cluster from Figure \ref{fig:kmeans}.}
\end{figure}

\begin{figure}[!htb]
    \centering
    \includegraphics[width=0.8\textwidth]{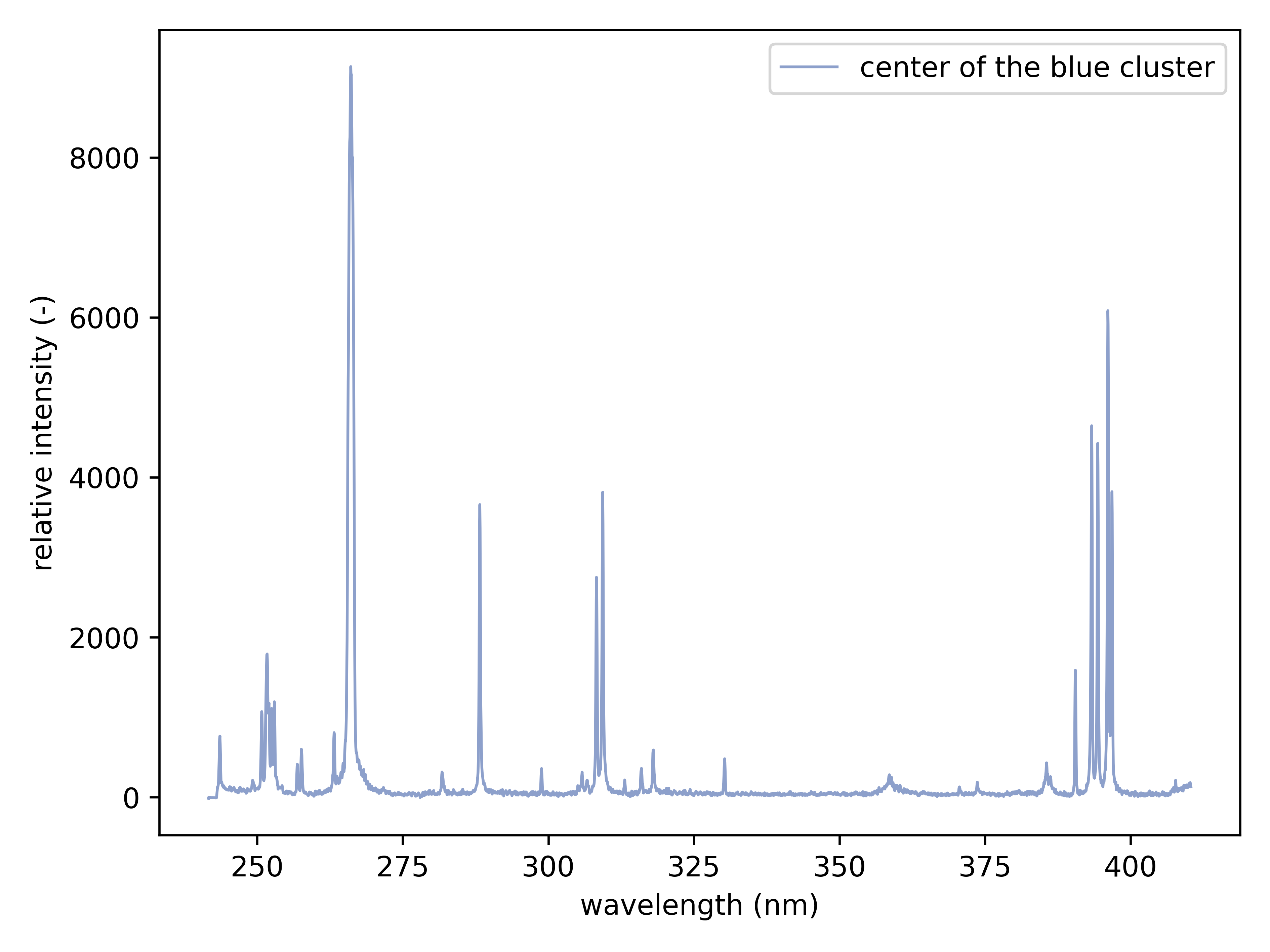}
    \caption{Center of the blue cluster from Figure \ref{fig:kmeans}.}
\end{figure}

\begin{figure}[!htb]
    \centering
    \includegraphics[width=0.8\textwidth]{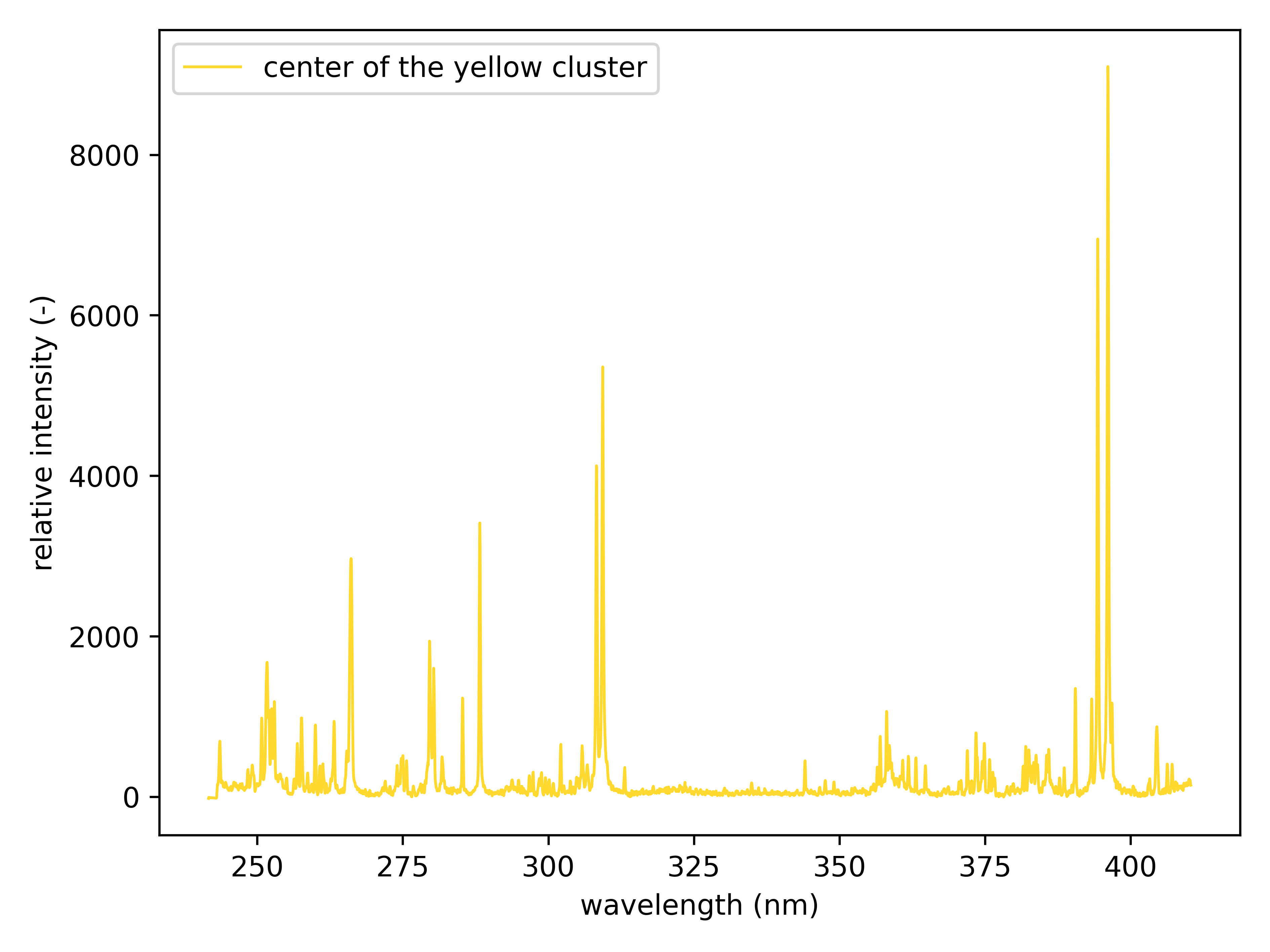}
    \caption{Center of the yellow cluster from Figure \ref{fig:kmeans}.}
\end{figure}

\begin{figure}[!htb]
    \centering
    \includegraphics[width=0.8\textwidth]{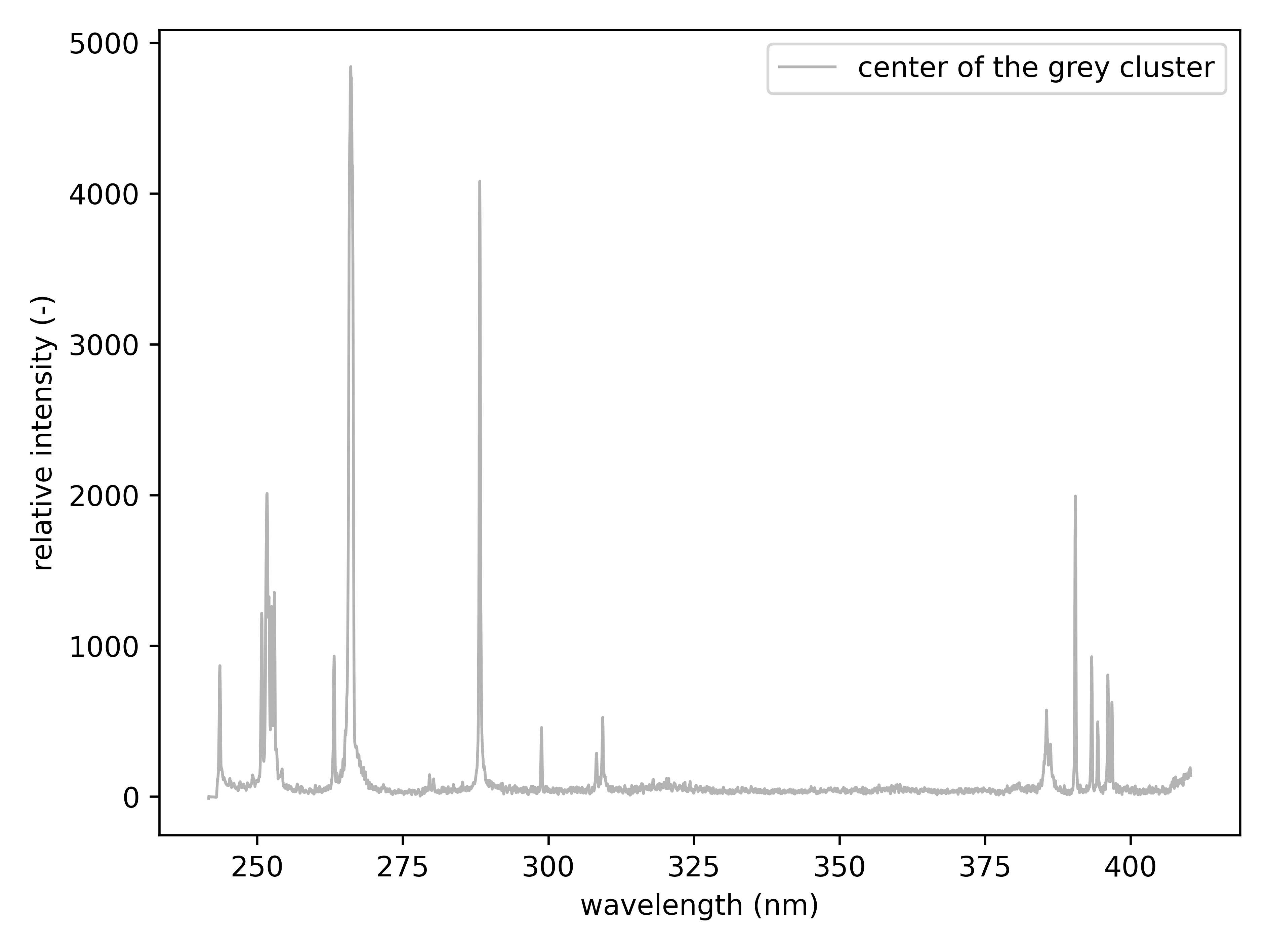}
    \caption{Center of the grey cluster from Figure \ref{fig:kmeans}.}
\end{figure}

\begin{figure}[!htb]
    \centering
    \includegraphics[width=0.8\textwidth]{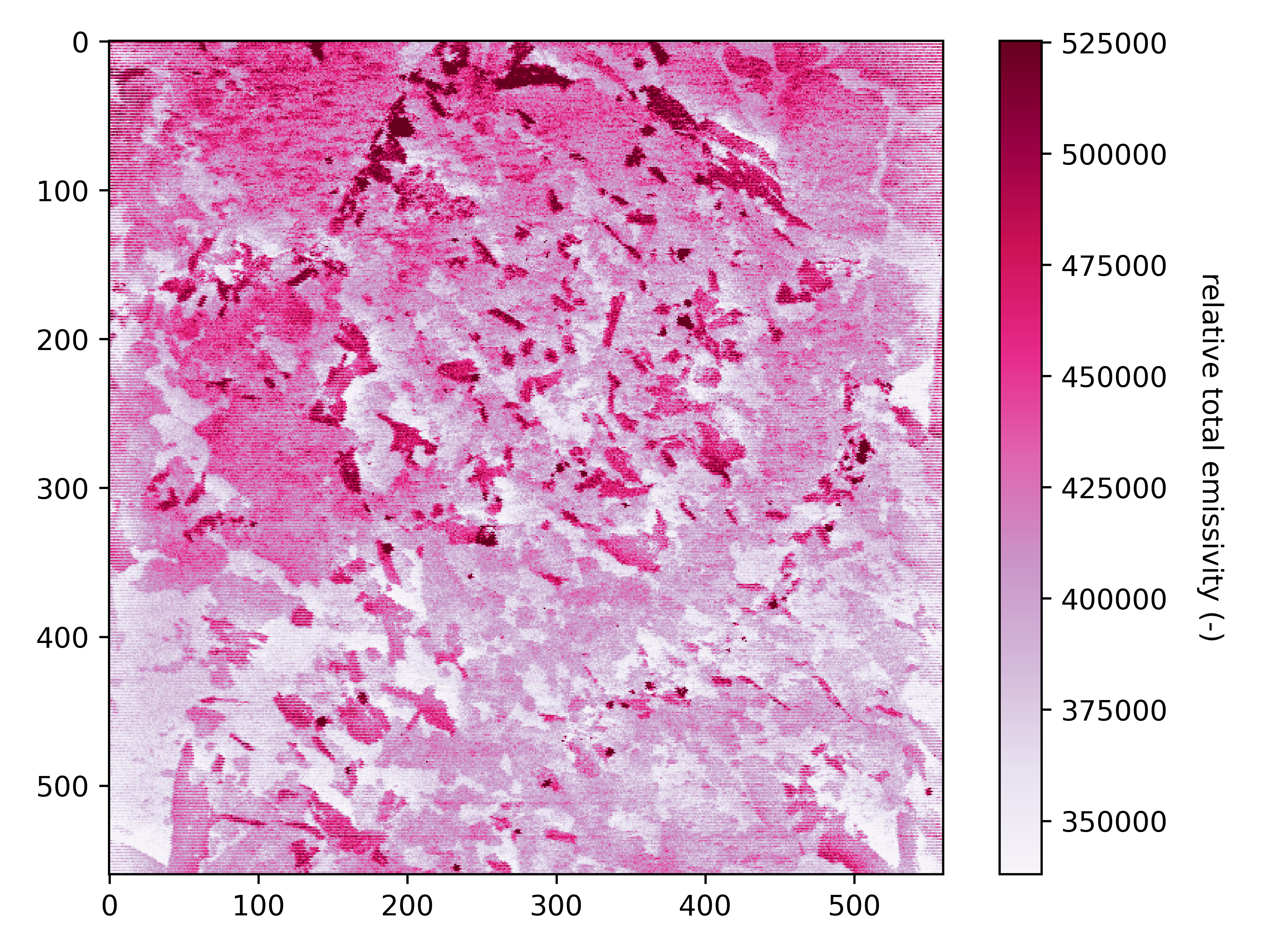}
    \caption{Total emissivity of the Primary training dataset.}
\end{figure}

\begin{figure}[!htb]
    \centering
    \includegraphics[width=0.8\textwidth]{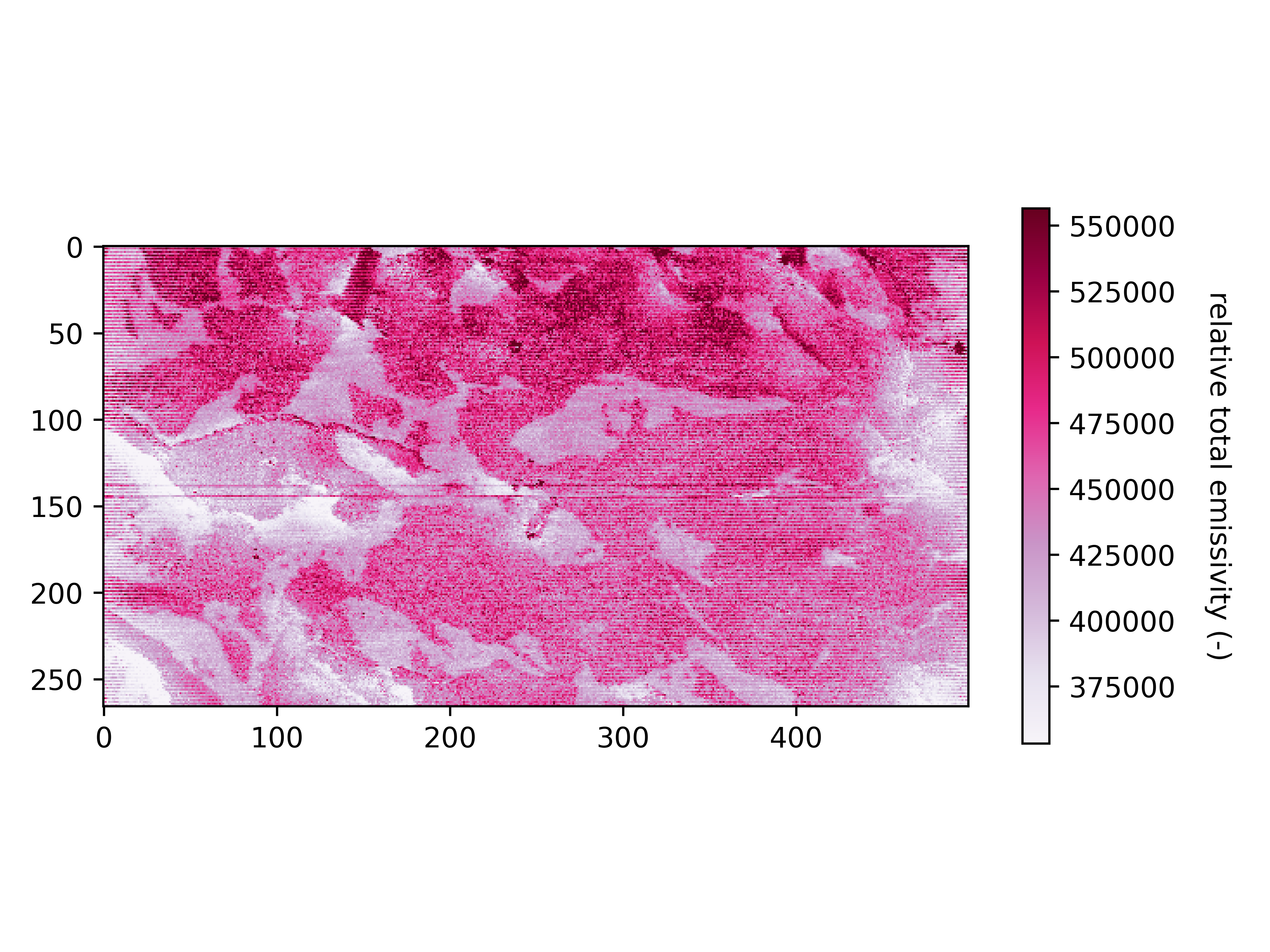}
    \caption{Total emissivity of the Primary validation dataset.}
\end{figure}

\begin{figure}[!htb]
    \centering
    \includegraphics[width=0.8\textwidth]{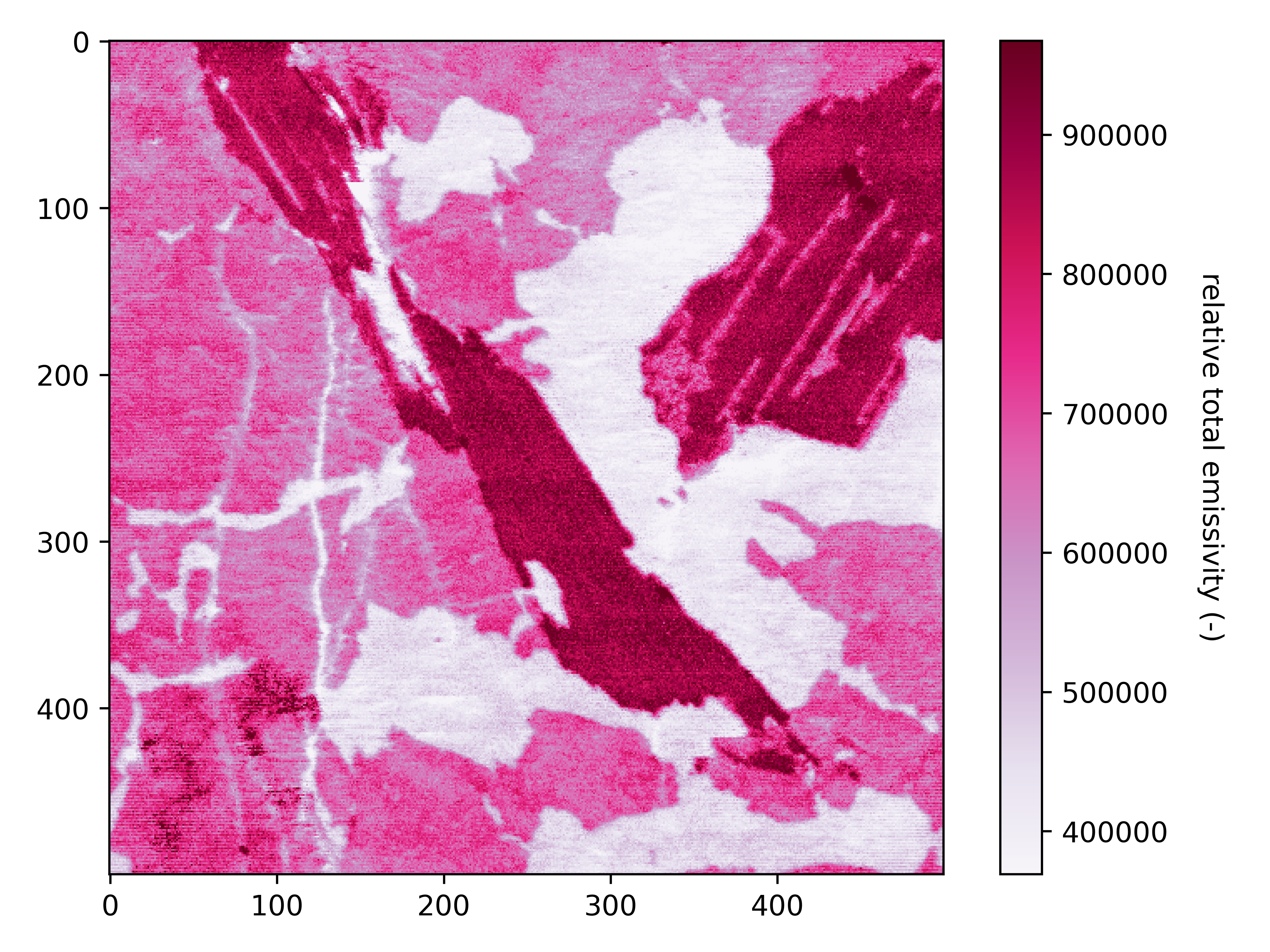}
    \caption{Total emissivity of the Primary test dataset.}
\end{figure}

\begin{figure}[!htb]
    \centering
    \includegraphics[width=0.8\textwidth]{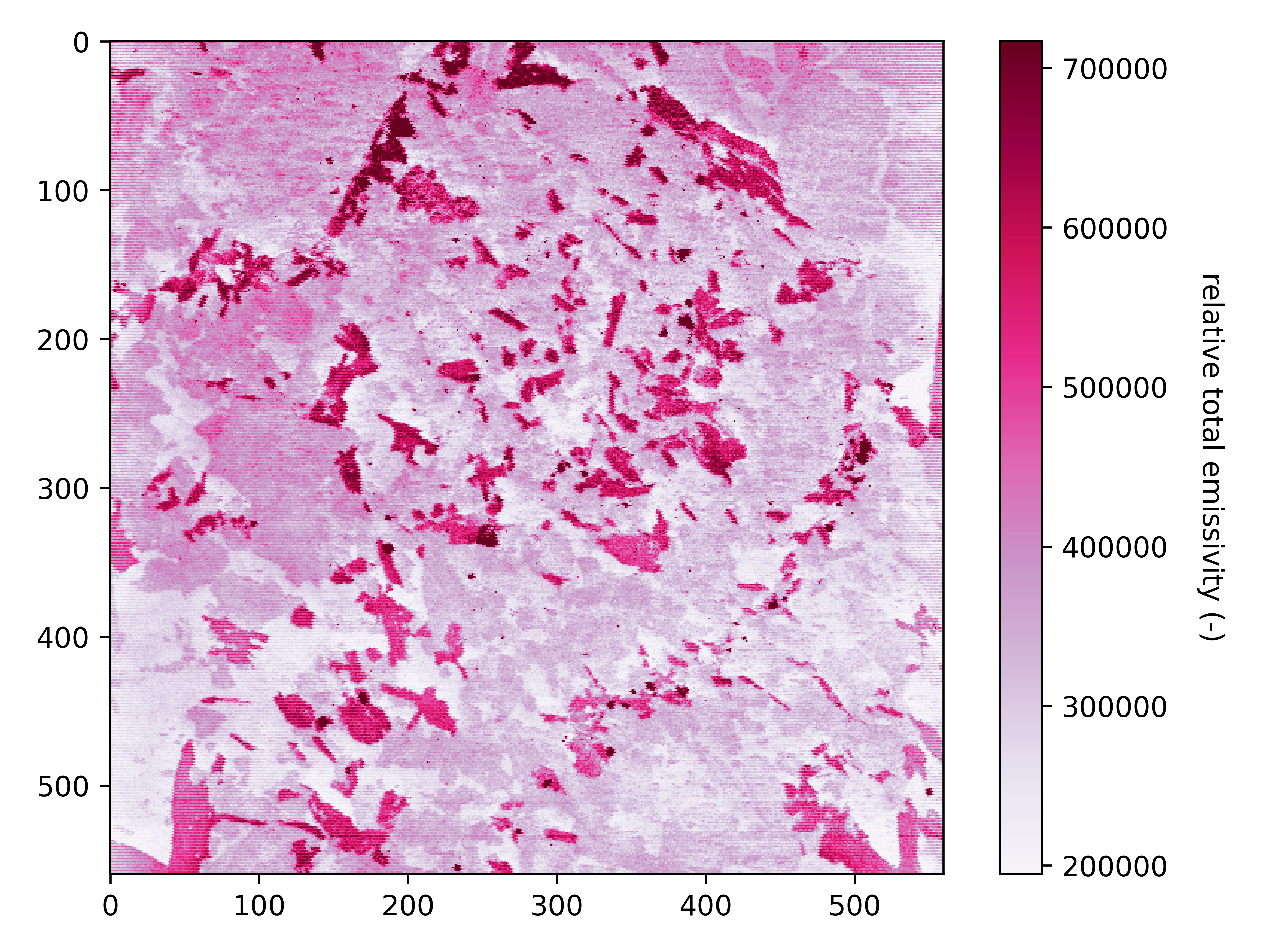}
    \caption{Total emissivity of the Secondary training dataset.}
\end{figure}

\begin{figure}[!htb]
    \centering
    \includegraphics[width=0.8\textwidth]{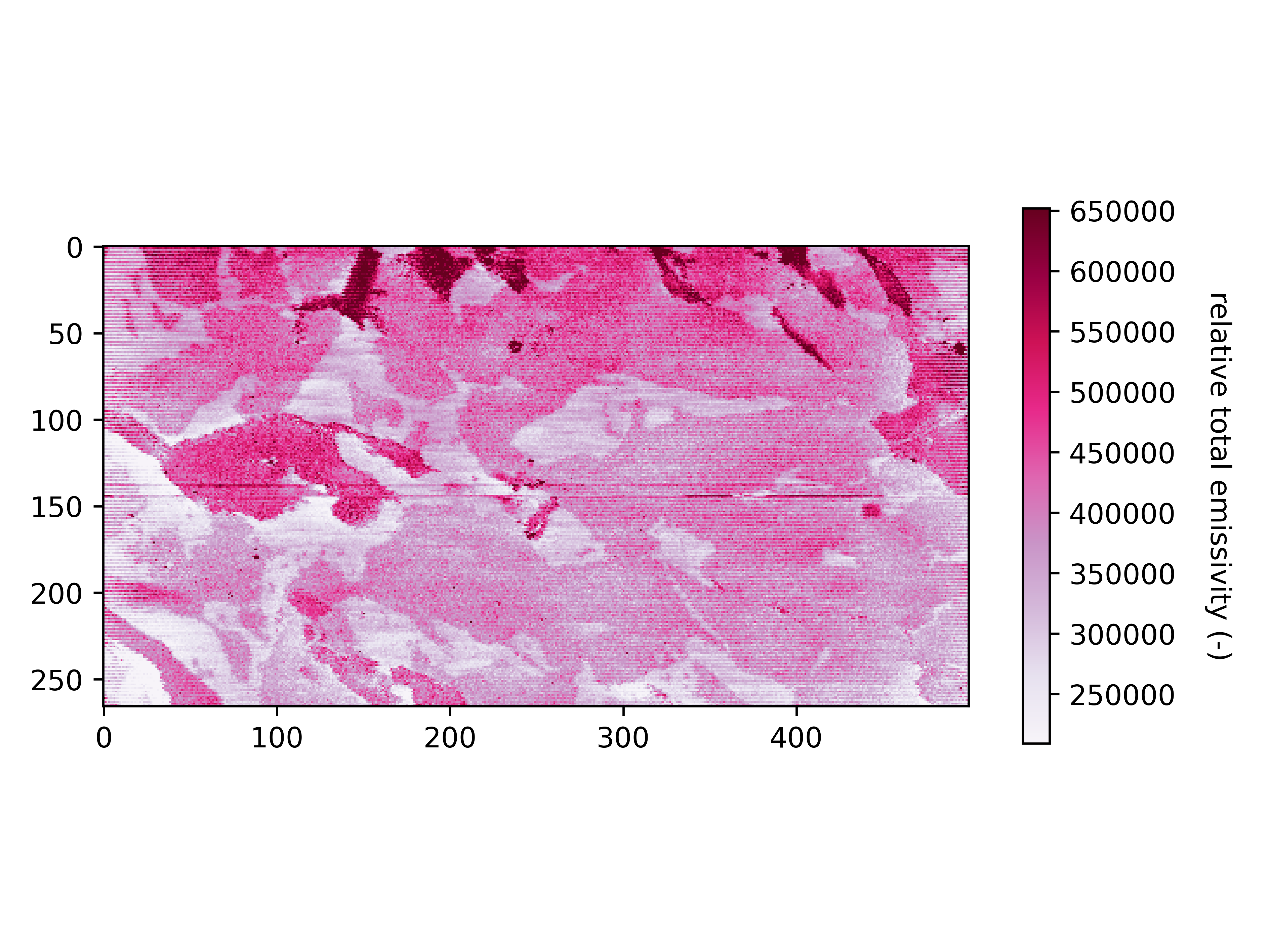}
    \caption{Total emissivity of the Secondary validation dataset.}
\end{figure}

\begin{figure}[!htb]
    \centering
    \includegraphics[width=0.8\textwidth]{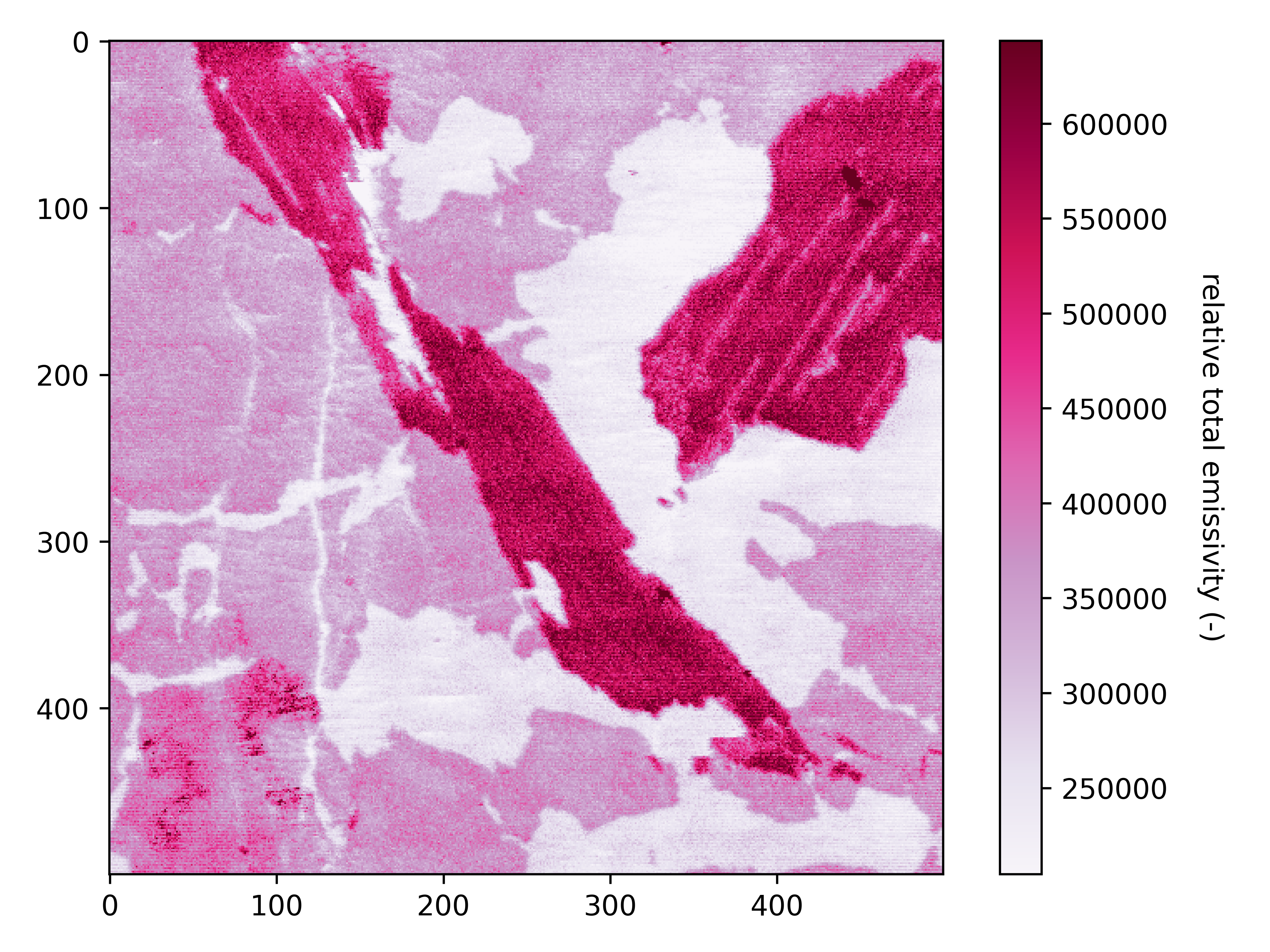}
    \caption{Total emissivity of the Secondary test dataset.}
\end{figure}

\begin{figure}[!htb]
    \centering
    \includegraphics[width=0.8\textwidth]{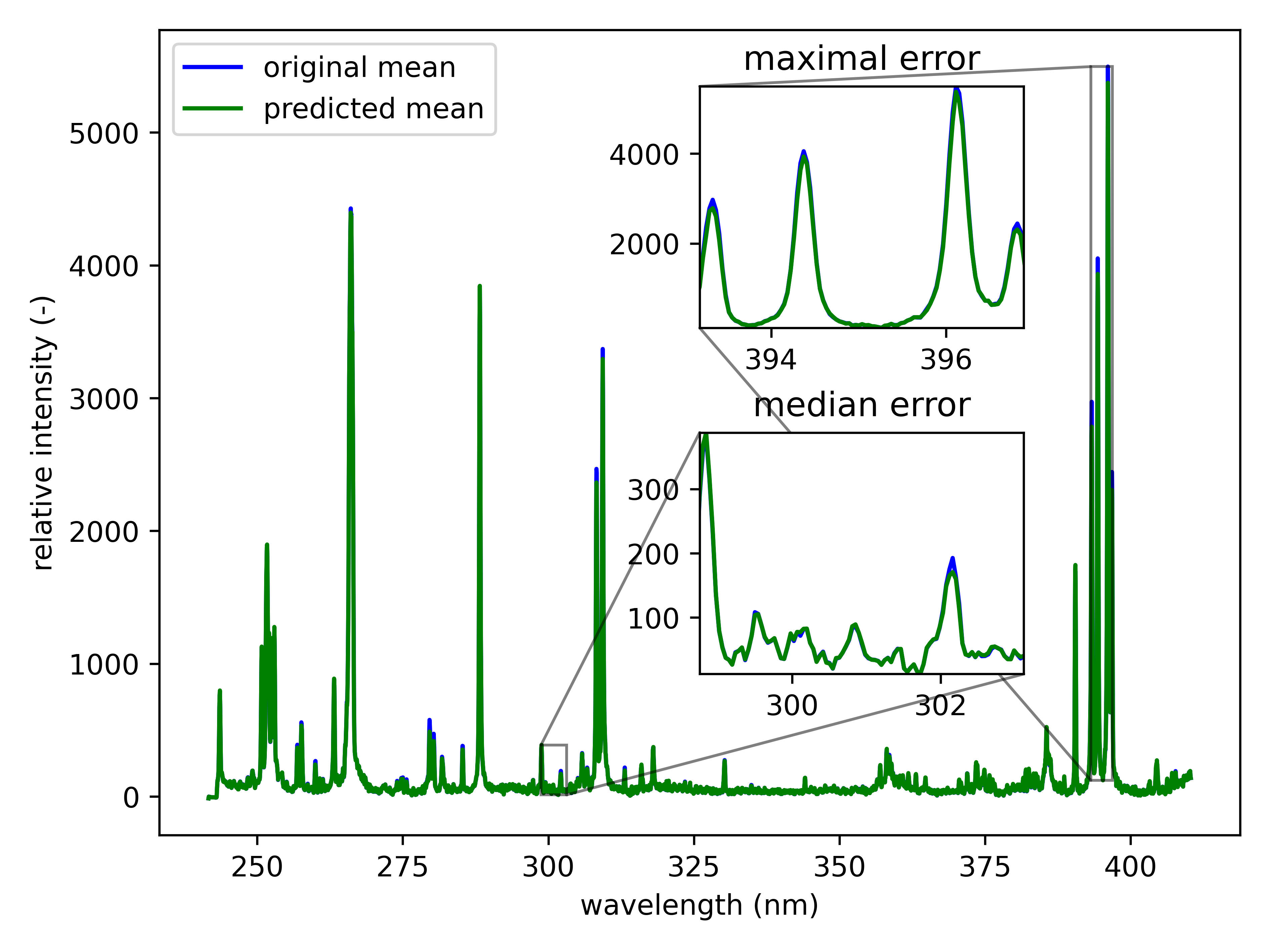}
    \caption{Comparison of the mean spectra from the $X_{Primary}$ (original) and $X_{Primary}^{\prime}$ (predicted) datasets.}
\end{figure}

\begin{figure}[!htb]
    \centering
    \includegraphics[width=0.8\textwidth]{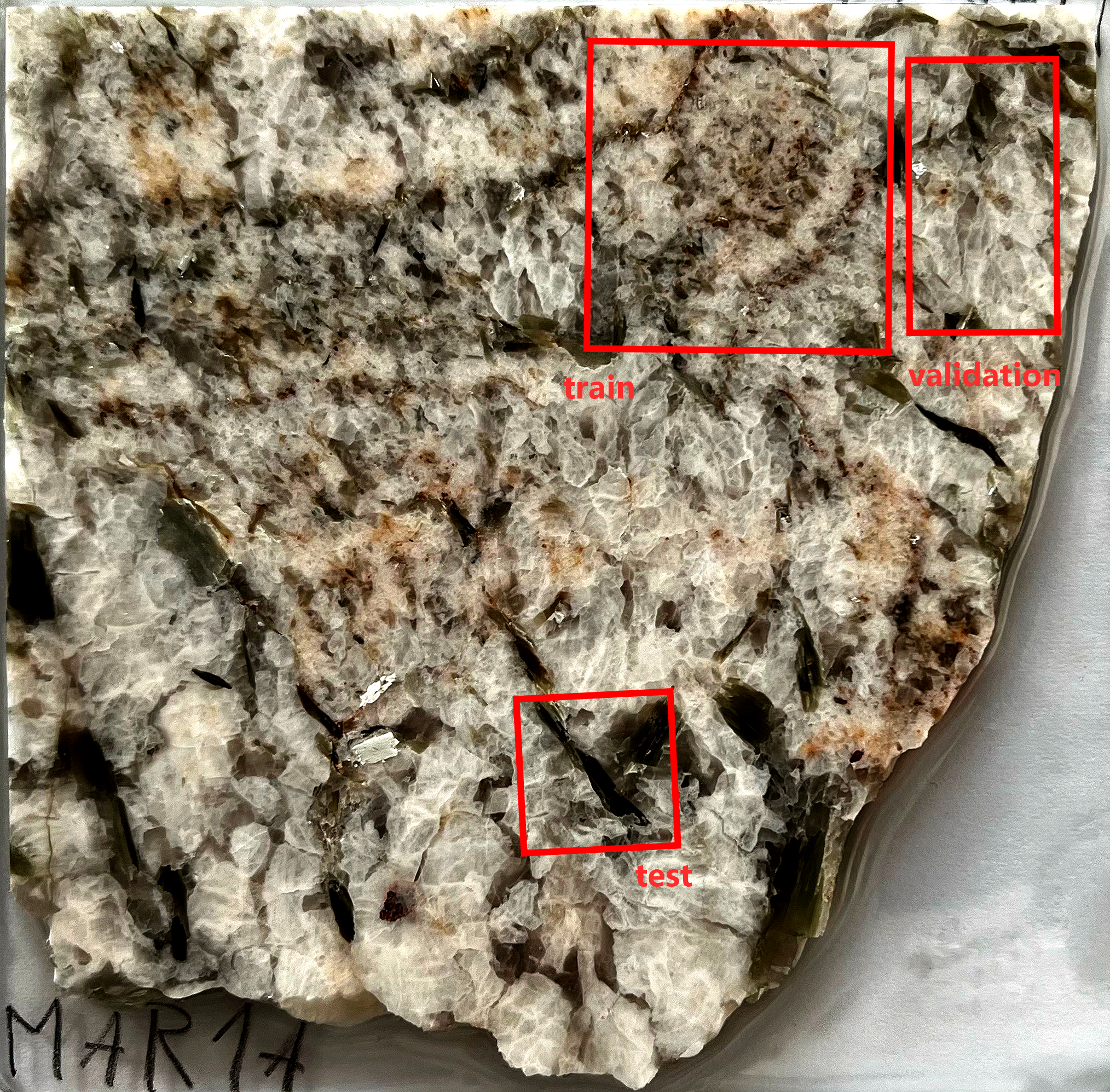}
    \caption{Photograph of the measured mineral sample.}
\end{figure}






\end{document}